\documentclass[letterpaper,english,twocolumn,amsmath,amssymb,superscriptaddress]{revtex4}
\usepackage[T1]{fontenc}
\usepackage{babel}
\usepackage{prettyref}
\usepackage{textcomp}
\usepackage{mathtools}
\usepackage{bm}
\usepackage{amsmath}
\usepackage{amssymb}
\usepackage{graphicx}
\usepackage[unicode=true,
 bookmarks=true,bookmarksnumbered=false,bookmarksopen=false,
 breaklinks=false,pdfborder={0 0 1},backref=false,colorlinks=false]
 {hyperref}
\hypersetup{pdftitle={A Microscopic Theory of Intrinsic Timescales in Spiking Neural Networks},
 pdfauthor={Alexander van Meegen, Sacha van Albada},
 colorlinks=true,linkcolor=black,citecolor=black,urlcolor=black,filecolor=black}

\makeatletter

\pdfpageheight\paperheight
\pdfpagewidth\paperwidth

\@ifundefined{textcolor}{}
{%
 \definecolor{BLACK}{gray}{0}
 \definecolor{WHITE}{gray}{1}
 \definecolor{RED}{rgb}{1,0,0}
 \definecolor{GREEN}{rgb}{0,1,0}
 \definecolor{BLUE}{rgb}{0,0,1}
 \definecolor{CYAN}{cmyk}{1,0,0,0}
 \definecolor{MAGENTA}{cmyk}{0,1,0,0}
 \definecolor{YELLOW}{cmyk}{0,0,1,0}
}

\usepackage[latin1]{inputenc}

\newrefformat{eq}{\hyperref[#1]{Eq.~(\ref{#1})}}
\newrefformat{eqs}{\hyperref[#1]{Eqs.~(\ref{#1})}}
\newrefformat{cap}{\hyperref[#1]{Fig.~\ref{#1}}}
\newrefformat{fig}{\hyperref[#1]{Fig.~\ref{#1}}}
\newrefformat{tab}{\hyperref[#1]{Table ~\ref{#1}}}
\newrefformat{sec}{\hyperref[#1]{Section~\ref{#1}}}
\newrefformat{subsec}{\hyperref[#1]{Subsection~\ref{#1}}}
\newrefformat{sub}{\hyperref[#1]{Section~\ref{#1}}}
\newrefformat{cha}{\hyperref[#1]{Chapter~\ref{#1}}}

\usepackage{MnSymbol}
\DeclareMathAlphabet\mathbb{U}{msb}{m}{n}

\makeatother

\begin{document}
\global\long\def\connmatrix{\mathcal{J}}%
\global\long\def\synmatrix{J}%
\global\long\def\sign{\operatorname{sign}}%
\global\long\def\erf{\operatorname{erf}}%
\global\long\def\probit{\operatorname{probit}}%

\title{A Microscopic Theory of Intrinsic Timescales in Spiking Neural Networks}
\author{Alexander van Meegen}
\affiliation{Institute of Neuroscience and Medicine (INM-6) and Institute for Advanced
Simulation (IAS-6) and JARA-Institut Brain Structure-Function Relationships
(INM-10), Jülich Research Centre, 52425 Jülich, Germany}
\affiliation{Institute of Zoology, University of Cologne, 50674 Cologne, Germany}
\author{Sacha J. van Albada}
\affiliation{Institute of Neuroscience and Medicine (INM-6) and Institute for Advanced
Simulation (IAS-6) and JARA-Institut Brain Structure-Function Relationships
(INM-10), Jülich Research Centre, 52425 Jülich, Germany}
\affiliation{Institute of Zoology, University of Cologne, 50674 Cologne, Germany}
\date{\today}
\begin{abstract}
A complex interplay of single-neuron properties and the recurrent
network structure shapes the activity of cortical neurons. The single-neuron
activity statistics differ in general from the respective population
statistics, including spectra and, correspondingly, autocorrelation
times. We develop a theory for self-consistent second-order single-neuron
statistics in block-structured sparse random networks of spiking neurons.
In particular, the theory predicts the neuron-level autocorrelation
times, also known as \emph{intrinsic timescales}, of the neuronal
activity. The theory is based on an extension of dynamic mean-field
theory from rate networks to spiking networks, which is validated
via simulations. It accounts for both static variability, e.g.~due
to a distributed number of incoming synapses per neuron, and temporal
fluctuations of the input. We apply the theory to balanced random
networks of generalized linear model neurons, balanced random networks
of leaky integrate-and-fire neurons, and a biologically constrained
network of leaky integrate-and-fire neurons. For the generalized linear
model network with an error function nonlinearity, a novel analytical
solution of the colored noise problem allows us to obtain self-consistent
firing rate distributions, single-neuron power spectra, and intrinsic
timescales. For the leaky integrate-and-fire networks, we derive an
approximate analytical solution of the colored noise problem, based
on the Stratonovich approximation of the Wiener-Rice series and a
novel analytical solution for the free upcrossing statistics. Again
closing the system self-consistently, in the fluctuation-driven regime
this approximation yields reliable estimates of the mean firing rate
and its variance across neurons, the inter-spike interval distribution,
the single-neuron power spectra, and intrinsic timescales. With the
help of our theory we find parameter regimes where the intrinsic timescale
significantly exceeds the membrane time constant, which indicates
the influence of the recurrent dynamics. Although the resulting intrinsic
timescales are on the same order for generalized linear model neurons
and leaky integrate-and-fire neurons, the two systems differ fundamentally:
For the former, the longer intrinsic timescale arises from an increased
firing probability after a spike; for the latter, it is a consequence
of a prolonged effective refractory period with a decreased firing
probability. Furthermore, the intrinsic timescale attains a maximum
at a critical synaptic strength for generalized linear model networks,
in contrast to the minimum found for leaky integrate-and-fire networks.
\end{abstract}
\maketitle

\section{Introduction}

Neural dynamics in the cerebral cortex of awake behaving animals unfolds
over multiple timescales, ranging from milliseconds up to seconds
and more \citep{Bernacchia11_366,Murray14,Honey12_423,Runyan17_92,Ogawa10_2433}.
Such a heterogeneity of timescales in the dynamics is a substrate
for temporal processing of sensory stimuli \citep{Mauk04_307} and
reflects integration of information over different time intervals
\citep{Honey12_423,Runyan17_92}. Intriguingly, \textit{in vivo} electrophysiological
recordings reveal a structure in the autocorrelation timescales of
the activity on the level of single neurons \citep{Murray14,RossiPool21_1}.
This structure could arise from systematic variations in single-neuron
or synaptic properties \citep{Duarte17_156,Wang20_169}, from the
intricate cortical network structure \citep{Chaudhuri2015_419}, or
from a combination of both \citep{Huntenburg18_21,Goulas18_775}.
Furthermore, timescales may be influenced by the external input to
the network, and depend on the chosen measurement procedure \citep{Marom10_16}.
Thus, while these timescales are referred to as \emph{intrinsic timescales},
they are shaped by intrinsic and extrinsic factors alike.

Explaining the timescales of individual neurons embedded in a network
poses a theoretical challenge: How to account for a microscopic, neuron-level
observable in a macroscopic theory? Clearly, a straightforward coarse-graining
of the activity eliminates the microscopic observable of interest
\citep{Wilting19_2759}. \emph{Dynamic mean-field theory} (DMFT) \citep{Sompolinsky88_259,Crisanti18_062120,Helias20_970}
makes microscopic observables accessible because, instead of coarse-graining
the activity of the neurons, it coarse-grains their input. Here, the
term `dynamic' specifies that the input is approximated as a stochastic
process that varies in time, in contrast to the notion of a mean-field
theory in physics, which usually describes processes embedded in a
constant field. DMFT has led to significant insights into the interrelation
between network structure and intrinsic timescales for recurrent networks
of (non-spiking) rate neurons \citep{Sompolinsky88_259,Kadmon15_041030,Huang17_31,Mastrogiuseppe17_e1005498,Schuecker18_041029,Crisanti18_062120,Beiran19_e1006893,Muscinelli19_e1007122,Helias20_970}.
In particular, it has been shown that very slow intrinsic timescales
emerge close to a transition to chaos in autonomous networks \citep{Sompolinsky88_259}.
Interestingly, simply adding a noisy input to the network significantly
reduces this effect and even leads to a novel dynamical regime \citep{Schuecker18_041029}.
Furthermore, increasing the complexity of the single-neuron dynamics
reveals that timescales of slow adaptive currents are not straightforwardly
expressed in the network dynamics \citep{Beiran19_e1006893}, and
leads to yet another dynamical regime termed ``resonant chaos''
\citep{Muscinelli19_e1007122}. In combination, these results suggest
that the mechanisms shaping the intrinsic timescales in recurrent
networks are highly involved.

A characteristic feature of neural communication in the brain is the
spike-based coupling \citep{Gerstner14}. Consequently, spiking neural
network models have already yielded notable insights into cortical
neural dynamics. Prominent examples are the excitatory-inhibitory
balance mechanism which dynamically generates strong fluctuations
while keeping the activity in a physiological range \citep{Amit97,Brunel00}
and the mechanism of recurrent inhibitory feedback leading to low
cross-correlation between neurons despite the high number of shared
inputs \citep{Tetzlaff12_e1002596,Helias14}. From a theoretical perspective,
spike-based coupling further increases the complexity of the dynamics.
This calls for an extension of DMFT to spiking networks, as recently
achieved with a model-independent framework \citep{Keup21_021064}
(see also \citep{FulviMari00}).

Perhaps unintuitively, the main obstacle is not the reduction of the
recurrent dynamics to the DMFT but the colored noise problem: to obtain
the output statistics of the neuron for temporally correlated input
statistics. Previous works relied on numerical methods to address
the colored noise problem \citep{Lerchner06_131,Lerchner06_634,dummer14,Pena18_9}
because the spiking nonlinearity renders this problem in general analytically
intractable. Such a self-consistent numerical scheme already revealed
an unexpected minimum instead of a maximum in the intrinsic timescales
for spiking networks at a critical coupling strength \citep{Wieland2015_040901}.
However, numerical solutions have the drawback that they lead to noisy
estimates of the autocorrelation function, which poses additional
challenges on the inference of intrinsic timescales \citep{Zeraati20_bioRxiv}
and other dynamical quantities from the neuronal and network parameters.
In addition, such a self-consistent numerical scheme is computationally
intensive.

In this paper, we use analytical approaches to close the self-consistency
equations for spiking networks. First, we transfer the theory for
rate networks to one for spiking networks starting from the characteristic
functional of the recurrent input. This shows that the first two cumulants
(mean and variance) of the connectivity matrix suffice to fully characterize
the effective stochastic input, and automatically take the static
variabilities (firing rate, indegree) in the network into account.
Since it is based on DMFT, the resulting theory indeed accounts for
the timescales on the microscopic level, orthogonal to approaches
where the activity of a population of neurons is reduced to an effective
mesoscopic description \citep{Schwalger17_e1005507}. Second, we derive
an analytical solution to the colored noise problem for \emph{generalized
linear model} (GLM) neurons with exponential and error function nonlinearity.
Using these analytical solutions, we validate that the self-consistent
DMFT captures both the static second-order statistics, the distribution
of firing rates across neurons, and the dynamic second-order statistics,
the population-averaged autocorrelation function. Furthermore, we
use the theory to investigate the conditions for longer intrinsic
timescales, like those observed in \textit{in vivo} electrophysiological
recordings \citep{Murray14,RossiPool21_1}, in a balanced random network
of GLM neurons. Due to the analytical tractability, our theory exposes
the factors that shape the intrinsic timescale. Third, we derive a
numerically efficient analytical approximation for the colored noise
problem for \emph{leaky integrate-and-fire} (LIF) neurons in the noise-driven
regime based on the Wiener-Rice series and the Stratonovich approximation
thereof \citep{Stratonovich67,Verechtchaguina06_031108}. For a different
approach based on a Markovian embedding, which leads to multidimensional
Fokker-Planck equations with involved boundary conditions that are
solved numerically, see \citep{Vellmer19_023024}. In contrast,
our approximation leads to integrals of which the computationally
most involved ones can be solved analytically. Lastly, we use these
results to explore the parameter space of a balanced random network
of LIF neurons for long timescales, and apply the theory to a more
elaborate model with population--specific connection probabilities
that are constrained by biological data \citep{Potjans14_785}.

We start this manuscript with the derivation of the DMFT equations
from the characteristic functional of the recurrent input. The remainder
of the results is structured according to the neuron model: First
we consider GLM neurons with exponential and error function nonlinearity,
respectively, then we turn to LIF neurons. For each neuron model,
we begin by deriving the solution or approximation of the colored
noise problem. We then describe the numerical method to solve the
self-consistent DMFT equations for the given neuron type (GLM or LIF).
Subsequently, we use our theory to investigate the timescale in the
respective network models.

\section{Microscopic Theory of Intrinsic Timescales}

\begin{figure}
\includegraphics{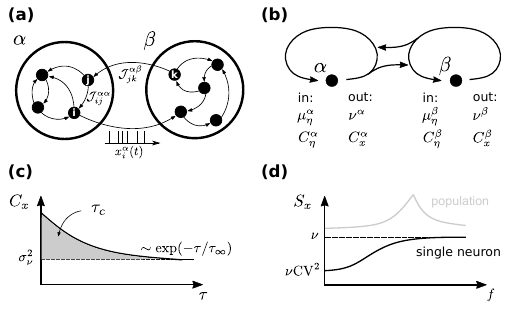}

\caption{Illustration of the theory. \textbf{(a)} We consider populations of
randomly connected neurons ($\alpha$, $\beta$) that communicate
via spike trains $x_{i}^{\alpha}(t)$. The neurons of population $\beta$
are connected to those of population $\alpha$ with connection probability
$p^{\alpha\beta}$. \textbf{(b)} The theory reduces a population to
a single neuron driven by an effective stochastic input $\eta^{\alpha}$.
The first- and second-order statistics $\mu_{\eta}^{\alpha}$ and
$C_{\eta}^{\alpha}$ of $\eta^{\alpha}$ depend self--consistently
on the output statistics, $\nu^{\alpha}$ and $C_{x}^{\alpha}$. \textbf{(c)}
From the stationary spike train autocorrelation function $\mbox{\ensuremath{C_{x}(\tau)=\nu\delta(\tau)+\hat{C}_{x}(\tau)}}$,
we obtain the correlation time $\tau_{c}$, the asymptotic decay $\tau_{\infty}$,
and the variability of the rate across neurons, $\sigma_{\nu}^{2}$.
\textbf{(d)} Instead of the stationary autocorrelation function we
sometimes consider the power spectrum $S_{x}(f)$, which saturates
at the firing rate, $S_{x}(f)\protect\overset{f\to\infty}{\to}\nu$,
and, for a renewal process, has the zero-frequency limit $\mbox{\ensuremath{S_{x}(f)\protect\overset{f\to0}{\to}\nu\mathrm{CV}^{2}}}$.
Throughout, we consider the population--averaged single-unit statistics
(black curve) instead of the statistics of the population--averaged
activity (gray curve). \label{fig:Sketch}}
\end{figure}

We consider random network topologies where the entries of the matrix
$\boldsymbol{\connmatrix}$ containing the synaptic strengths, i.e.~the
amplitudes of evoked post-synaptic currents due to incoming spikes,
are independent and identically distributed (i.i.d.). A synapse from
neuron $j$ to neuron $i$ exists ($\connmatrix_{ij}$ is non-zero)
with probability $p$; each non-zero entry $\synmatrix_{ij}$ is independently
sampled from the distribution of synaptic strengths with mean $\mu_{\synmatrix}$
and variance $\sigma_{\synmatrix}^{2}<\infty$:
\begin{align}
\connmatrix_{ij} & =\begin{cases}
J_{ij} & \text{with probability }p.\\
0 & \text{with probability }1-p.
\end{cases}\label{eq:mathcal_J}
\end{align}
The connectivity is thus taken to be pairwise Bernoulli, yielding
maximally one synapse from a given presynaptic to a given postsynaptic
neuron. To account for Dale's law and further heterogeneities, we
subdivide the network into populations, e.g., all pyramidal cells
in cortical layer V, consisting of statistically identical neurons
and denote the population by a Greek superscript. Within this generalization,
the entries of $\boldsymbol{\connmatrix}$ are still i.i.d.~random
numbers for a given pair of populations $\alpha$, $\beta$, but $p^{\alpha\beta}$
and the distribution of $\synmatrix_{ij}^{\alpha\beta}$ can vary
for different pairs of populations (\prettyref{fig:Sketch}\textbf{a}).
For example, if $I$ denotes a population of inhibitory interneurons,
all $\synmatrix_{ij}^{\alpha I}$ are negative.

In this manuscript, we focus on the situation where the average number
of synapses per neuron, the indegree $K^{\alpha\beta}=p^{\alpha\beta}N^{\beta}$,
is large: $K^{\alpha\beta}\gg1$ due to a large number of presynaptic
neurons $N^{\beta}\gg1$ in combination with a moderate connection
probability $p^{\alpha\beta}$ on the order of $10$\%, in agreement
with the situation in cortical networks \citep{Braitenberg98}. In
line with the theory of balanced networks \citep{Vreeswijk96_1724},
we assume that neither single spikes are sufficient to cause firing
nor coherent input from all presynaptic neurons is necessary. Moreover,
we consider networks which are in an asynchronous irregular state
exhibited by cortical networks of awake, behaving animals \citep{Harris11_509}.

In the following, we first consider a single population for clarity
because the generalization to multiple populations is straightforward.

\subsection{Input statistics}

Dynamic mean-field theory reduces the dynamics of the recurrent network
to a set of self-consistent stochastic equations. Its core idea is
to approximate the recurrent input
\begin{align}
\eta_{i}(t) & =\sum_{j=1}^{N}\connmatrix_{ij}x_{j}(t)\label{eq:input}
\end{align}
by independent Gaussian processes. Throughout this manuscript, $x_{i}(t)$
denotes the spike train of neuron $i$. The sum in \prettyref{eq:input}
extends over all $N$ neurons, using that $\connmatrix_{ij}=0$ for
neurons that are not connected.

\subsubsection{Gaussian Process Approximation\label{subsec:Gaussian-Process-Approximation}}

Here, we sketch the derivation to expose necessary conditions for
the DMFT. For the full treatment of the problem, we refer to the model--independent
DMFT developed in \citep{Keup21_021064}, which is applicable to spiking
networks.

We start from the deterministic input \prettyref{eq:input} and derive
its approximation as independent Gaussian processes. To this end,
let us consider the characteristic functional of the recurrent input.
Because $\eta_{i}(t)$ is a deterministic quantity, its distribution
is a Dirac delta and its characteristic functional, defined by $\Phi_{\boldsymbol{\eta}}[\boldsymbol{u}(t)]=\langle\exp(i\int_{0}^{T}\boldsymbol{u}(t)^{\intercal}\boldsymbol{\eta}(t)dt)\rangle_{\boldsymbol{\eta}}$,
is \citep{Feynman10,Stratonovich67} (see also Appendix A, \prettyref{eq:appendix_charfctl_deterministic})
\begin{align}
\Phi_{\boldsymbol{\eta}}[\boldsymbol{u}(t)] & =\exp\left(i\int_{0}^{T}\sum_{i,j=1}^{N}u_{i}(t)\connmatrix_{ij}x_{j}(t)dt\right).\label{eq:charfct_input}
\end{align}
Now we assume that the dynamics of the system are, on a statistical
level, very similar for any given realization if the connectivity,
i.e.~we assume that the system is self--averaging. Thus, we can
consider the average across realizations of $\boldsymbol{\connmatrix}$,
\begin{align*}
\langle\Phi_{\boldsymbol{\eta}}[\boldsymbol{u}(t)]\rangle_{\boldsymbol{\connmatrix}} & \approx e^{i\langle\connmatrix\rangle\sum_{i,j=1}^{N}\int_{0}^{T}u_{i}(t)x_{j}(t)dt}\\
 & \times e^{-\frac{1}{2}\langle{\Delta\connmatrix}^{2}\rangle\sum_{i,j=1}^{N}\left(\int_{0}^{T}u_{i}(t)x_{j}(t)dt\right)^{2}},
\end{align*}
where we used the independence of the $\connmatrix_{ij}$, their characteristic
function $\langle\exp(ik_{ij}\connmatrix_{ij})\rangle_{\connmatrix_{ij}}=\exp(i\langle\connmatrix\rangle k_{ij}-\tfrac{1}{2}\langle{\Delta\connmatrix}^{2}\rangle k_{ij}^{2}+\dots)$,
and neglected the cumulants of $\connmatrix_{ij}$ beyond the second-order
cumulant (the variance) $\langle{\Delta\connmatrix}^{2}\rangle$.
Due to the independence of the $\connmatrix_{ij}$, the expectation
factorizes into a product $\prod_{i,j=1}^{N}$ which leads to the
sum $\sum_{i,j=1}^{N}$ in the exponent. Within each factor, the first
(second) cumulant leads to a linear (quadratic) term in the exponent.
Next, we rewrite the square, $\left(\int_{0}^{T}u_{i}(t)x_{j}(t)dt\right)^{2}=\int_{0}^{T}\int_{0}^{T}u_{i}(t)u_{i}(t^{\prime})x_{j}(t)x_{j}(t^{\prime})dtdt^{\prime}$,
and introduce the auxiliary fields
\begin{align}
\mu_{\eta}(t) & =\langle\connmatrix\rangle\sum_{j=1}^{N}x_{j}(t),\label{eq:DMFT_aux_field_mu}\\
C_{\eta}(t,t^{\prime}) & =\langle{\Delta\connmatrix}^{2}\rangle\sum_{j=1}^{N}x_{j}(t)x_{j}(t^{\prime}).\label{eq:DMFT_aux_field_C}
\end{align}
Using the auxiliary fields, the characteristic functional factorizes,
$\langle\Phi_{\boldsymbol{\eta}}[\boldsymbol{u}(t)]\rangle_{\boldsymbol{\connmatrix}}\approx\prod_{i=1}^{N}\hat{\Phi}_{\eta}[u_{i}(t)]$,
with the individual factors given by
\begin{align*}
\hat{\Phi}_{\eta}[u(t)] & =e^{i\int_{0}^{T}u(t)\mu_{\eta}(t)dt-\frac{1}{2}\int_{0}^{T}\int_{0}^{T}u(t)C_{\eta}(t,t^{\prime})u(t^{\prime})dtdt^{\prime}},
\end{align*}
which is the characteristic functional of a Gaussian process with
mean $\mu_{\eta}(t)$ and correlation function $C_{\eta}(t,t^{\prime})$
\citep{Feynman10,Stratonovich67} (see Appendix A, \prettyref{eq:appendix_charfctl_GP}).
The factorization $\langle\Phi_{\boldsymbol{\eta}}[\boldsymbol{u}(t)]\rangle_{\boldsymbol{\connmatrix}}\approx\prod_{i=1}^{N}\hat{\Phi}_{\eta}[u_{i}(t)]$
implies that the approximate inputs described by $\hat{\Phi}_{\eta}[u(t)]$
are independent across neurons.

The above expressions reveal multiple assumptions we make in the DMFT.
First, we assumed self--averaging. This is a necessary assumption
if one wants to derive a statement that generalizes beyond a given
connectivity matrix to its statistics only. For a broad class of rate
networks, one can show rigorously that network--averaged quantities
are indeed self--averaging \citep{Ben-Arous95_455,vanMeegen20_2009}.
Here, we check this assumption post--hoc by comparison of the theory
with simulations for a single realization of the connectivity. Second,
we implicitly assumed $\bar{g}:=N\langle\connmatrix\rangle$ and $g^{2}:=N\langle{\Delta\connmatrix}^{2}\rangle$
do not scale with $N$ such that the auxiliary fields remain finite
for large networks. Using the mean number of inputs per neuron $K=pN$
and the properties of $\boldsymbol{\connmatrix}$, we get
\begin{align}
\bar{g}=K\mu_{J}, & \qquad g^{2}=K\left(\sigma_{J}^{2}+(1-p)\mu_{J}^{2}\right).\label{eqs:g_g_bar}
\end{align}
Third, we neglected higher cumulants of the input. Using the assumption
$J_{ij}=O(1/\sqrt{K})$ leads to $\mu_{J}=O(1/\sqrt{K})$, $\sigma_{J}^{2}=O(1/K)$
and thus $\bar{g}=O(\sqrt{K})$, $g^{2}=O(1)$ as well as $O(1/\sqrt{K})$
for the neglected higher cumulants. Accordingly, in the regime $K\gg1$,
neglecting the contributions from higher cumulants, e.g.~due to shot
noise effects \citep{dummer14}, is justified.

\subsubsection{Self--consistency problem}

Given these assumptions, the recurrent inputs $\eta_{i}(t)$ can be
approximated by independent Gaussian processes, which leads to a coarse-grained
description of the dynamics: since all inputs are statistically equivalent,
the neurons become statistically equivalent as well and the system
reduces to $N$ independent, identical stochastic equations. For $N\gg1$,
we can replace the empirical averages in \prettyref{eq:DMFT_aux_field_mu}
and \prettyref{eq:DMFT_aux_field_C} by ensemble averages such that
we arrive at a set of self-consistency equations. This step can be
made rigorous using the formalism of \citep{Keup21_021064}, see Eqs.~(2-3)
and Appendix 1 therein.

In the stationary state, the self-consistency equations are given
by
\begin{align}
\mu_{\eta}=\bar{g}\,\langle x\rangle_{\eta}, & \qquad C_{\eta}(\tau)=g^{2}\,\langle xx\rangle_{\eta}(\tau).\label{eqs:DMFT_1pop}
\end{align}
The averages $\langle x\rangle_{\eta}\equiv\nu$ and $\langle xx\rangle_{\eta}(\tau)-\nu^{2}\equiv C_{x}(\tau)$
denote the mean (firing rate) and correlation function of the spike
train produced by a neuron driven by the effective stochastic input
$\eta(t)$. Since the input thereby appears on both the left-hand
and the right-hand side, this poses a self-consistency problem. To
recapitulate, DMFT approximates the input of a single neuron by an
effective Gaussian process with self-consistent statistics (\prettyref{fig:Sketch}\textbf{b}).
Thus, the description, albeit stochastic, is still on the level of
individual neurons.

\subsubsection{Static contribution}

The networks we consider are heterogeneous even within a population---each
neuron potentially has a different number of presynaptic partners
and thus also a different firing rate \citep{Roxin11_16217}. On a
first glance, DMFT neglects this heterogeneity. However, \prettyref{eqs:DMFT_1pop}
in fact account for such static variabilities: on the r.h.s.~the
second moment of the spike train appears instead of the correlation
function. Rewriting $\langle xx\rangle_{\eta}(\tau)=C_{x}(\tau)+\nu^{2}$
reveals a first static component $g^{2}\nu^{2}$ of the variability
of the effective input due to the firing rate of individual neurons.
Moreover, $C_{x}(\tau\to\infty)\equiv\sigma_{\nu}^{2}$ potentially
saturates on a plateau which accounts for the variability of the firing
rate across neurons (\prettyref{fig:Sketch}\textbf{c}). To make this
explicit, we sometimes rewrite 
\begin{align}
\eta(t) & =\zeta+\xi(t),\label{eq:DMFT_static}
\end{align}
where $\zeta$ is a Gaussian random variable with $\mu_{\zeta}=\bar{g}\nu$,
$\sigma_{\zeta}^{2}=g^{2}(\nu^{2}+\sigma_{\nu}^{2})$ and $\xi(t)$
a zero-mean Gaussian process with $C_{\xi}(\tau)=g^{2}(C_{x}(\tau)-\sigma_{\nu}^{2})$.

\subsection{Multiple populations}

Using the expressions \prettyref{eqs:DMFT_1pop} for a single population,
we can straightforwardly generalize the theory to multiple populations.
Due to the independence of the effective inputs in DMFT, both mean
and correlation function are a simple sum over the contributions from
all populations \citep{Kadmon15_041030,Aljadeff15_088101}:
\begin{align}
\mu_{\eta}^{\alpha} & =\sum_{\beta}\bar{g}^{\alpha\beta}\nu^{\beta},\label{eq:DMFT_mean_pop}\\
C_{\eta}^{\alpha}(\tau) & =\sum_{\beta}{(g^{\alpha\beta})}^{2}\left(C_{x}^{\beta}(\tau)+{(\nu^{\beta})}^{2}\right),\label{eq:DMFT_correlation_pop}
\end{align}
with the corresponding generalizations of \prettyref{eqs:g_g_bar},
$\mbox{\ensuremath{\bar{g}^{\alpha\beta}=K^{\alpha\beta}\mu_{\synmatrix}^{\alpha\beta}}}$
and $\mbox{\ensuremath{{(g^{\alpha\beta})}^{2}=K^{\alpha\beta}\left({(\sigma_{\synmatrix}^{\alpha\beta})}^{2}+(1-p^{\alpha\beta}){(\mu_{\synmatrix}^{\alpha\beta})}^{2}\right)}}$.
This leads to one stochastic equation per population (\prettyref{fig:Sketch}\textbf{b}).
As before, we can split the input into static and dynamic contributions,
$\eta^{\alpha}(t)=\zeta^{\alpha}+\xi^{\alpha}(t)$.

\subsubsection{External input}

We take the sum $\sum_{\beta}$ to include external populations, e.g.~excitatory
neurons that drive the network dynamics with homogeneous Poissonian
spike trains of rate $\nu^{\mathrm{ext}}$. In \prettyref{eq:DMFT_mean_pop}
and \prettyref{eq:DMFT_correlation_pop}, such an external Poisson
input leads to a term $J^{\alpha,\mathrm{ext}}\nu^{\mathrm{ext}}$
and ${(J^{\alpha,\mathrm{ext}})}^{2}\nu^{\mathrm{ext}}\delta(\tau)$,
respectively. If the network is driven by a constant external input,
only \prettyref{eq:DMFT_mean_pop} obtains an additional contribution
$\mu_{\mathrm{ext}}^{\alpha}$. An external zero-mean, stationary
Gaussian process leads to an additional term $C_{\mathrm{ext}}(\tau)$
in \prettyref{eq:DMFT_correlation_pop}.

\subsection{Output statistics}

Approximating the input is only the first step. In a second step,
the self-consistency problem has to be solved. To this end, the output
statistics of a neuron driven by a non-Markovian Gaussian process
have to be calculated. In other words, we need a solution for the
colored noise problem. The full non-Markovian problem has to be considered
because a Markovian approximation neglects the quantity of interest:
the temporal correlations. For sufficiently simple rate neurons, the
problem is analytically solvable \citep{Sompolinsky88_259,vanMeegen18_258302};
the case of two spiking neuron models is discussed in the following
sections. For the remainder of this section, let us assume that we
are able to solve the colored noise problem to obtain a self--consistent
solution of \prettyref{eq:DMFT_mean_pop} and \prettyref{eq:DMFT_correlation_pop}.

\subsubsection{Timescale}

Given a self-consistent solution, we can calculate the intrinsic timescale
from the spike-train autocorrelation function $C_{x}^{\alpha}(\tau)$.
Since $C_{x}^{\alpha}(\tau)$ always contains a delta peak \citep{Stratonovich67},
we consider only the smooth part of the autocorrelation function $\hat{C}_{x}^{\alpha}(\tau)\equiv C_{x}^{\alpha}(\tau)-\nu^{\alpha}\delta(\tau)$.
To characterize the timescale, we use the definition of \citep{Stratonovich67}
(\prettyref{fig:Sketch}\textbf{c}):
\begin{align}
\tau_{c}^{\alpha} & =\int_{0}^{\infty}\left|\frac{\hat{C}_{x}^{\alpha}(\tau)-\hat{C}_{x}^{\alpha}(\infty)}{\hat{C}_{x}^{\alpha}(0)-\hat{C}_{x}^{\alpha}(\infty)}\right|d\tau.\label{eq:def_taucorr}
\end{align}
Note that the definition of the autocorrelation time is not unequivocal.
Other possible definitions include $\mbox{\ensuremath{\tau_{c}^{\alpha}=\int_{-\infty}^{\infty}\left|\frac{\hat{C}_{x}^{\alpha}(\tau)-\hat{C}_{x}^{\alpha}(\infty)}{\hat{C}_{x}^{\alpha}(0)-\hat{C}_{x}^{\alpha}(\infty)}\right|^{2}d\tau}}$
\citep{Wieland2015_040901} and $\mbox{\ensuremath{\tau_{c}^{\alpha}=\frac{\int_{0}^{\infty}\tau|\hat{C}_{x}^{\alpha}(\tau)-\hat{C}_{x}^{\alpha}(\infty)|d\tau}{\int_{0}^{\infty}|\hat{C}_{x}^{\alpha}(\tau)-\hat{C}_{x}^{\alpha}(\infty)|d\tau}}}$
\citep{Muscinelli19_e1007122}. We observed drastic differences between
these definitions for empirical correlation functions directly obtained
from the simulations. These differences are in part an artifact from
the absolute value: the variance of the empirical estimate grows with
$\tau$ \citep{Grigera20_arXiv}; due to the absolute value these
fluctuations add up. The three functional forms carry with them different
fluctuations, e.g.~the squared fluctuations $\left|\frac{\hat{C}_{x}^{\alpha}(\tau)-\hat{C}_{x}^{\alpha}(\infty)}{\hat{C}_{x}^{\alpha}(0)-\hat{C}_{x}^{\alpha}(\infty)}\right|^{2}$
are typically much smaller than $\left|\frac{\hat{C}_{x}^{\alpha}(\tau)-\hat{C}_{x}^{\alpha}(\infty)}{\hat{C}_{x}^{\alpha}(0)-\hat{C}_{x}^{\alpha}(\infty)}\right|<1$,
and hence lead to different estimates. For theoretically predicted
autocorrelations, the difference is less drastic and we choose \prettyref{eq:def_taucorr}
because it is the most simple definition. Due to this difficulty,
we always use the theoretical prediction of the autocorrelation function
to determine the timescale---after checking that it matches the empirical
autocorrelation function well apart from fluctuations.

In addition to $\tau_{c}^{\alpha}$, we will also consider the asymptotic
decay constant (\prettyref{fig:Sketch}\textbf{c})
\begin{equation}
\hat{C}_{x}^{\alpha}(\tau)-\hat{C}_{x}^{\alpha}(\infty)\sim\exp(-\tau/\tau_{\infty}^{\alpha}),\label{eq:def_tauinf}
\end{equation}
because in special cases $\tau_{\infty}^{\alpha}$ directly follows
from our theory. For a simple exponential autocorrelation function,
the timescales in Eqs.~\eqref{eq:def_taucorr} and \eqref{eq:def_tauinf}
coincide. We work from the assumption that \prettyref{eq:def_tauinf}
is a good approximation to \prettyref{eq:def_taucorr} and verify
this assumption post-hoc.

Recently, two approaches have been proposed to estimate the timescale
directly from spiking data \citep{Wilting18_2325,Zeraati20_bioRxiv}.
While both overcome important challenges, biases in the estimated
timescale related to and independent of subsampling, respectively,
we do not use them here because they rely on models which implicitly
assume (a mixture of) exponential correlation functions: \citep{Wilting18_2325}
assumes an autoregressive model and \citep{Zeraati20_bioRxiv} a mixture
of Ornstein-Uhlenbeck processes.

\subsubsection{Spike train power spectrum}

Instead of the autocorrelation function, we sometimes consider the
spike train power spectrum (\prettyref{fig:Sketch}\textbf{d})
\begin{align}
S_{x}^{\alpha}(f) & =\int_{-\infty}^{\infty}e^{2\pi if\tau}C_{x}^{\alpha}(\tau)d\tau.\label{eq:powerspec}
\end{align}
Due to the delta peak in the autocorrelation function, the power spectrum
always saturates at the firing rate, $S_{x}^{\alpha}(f)\overset{f\to\infty}{\to}\nu^{\alpha}$.
For a renewal process, the zero-frequency limit is $S_{x}^{\alpha}(f)\overset{f\to0}{\to}\nu^{\alpha}\mathrm{CV}_{\alpha}^{2}$
\citep{Gerstner14}, which directly reveals the coefficient of variation
of the inter-spike-interval (ISI) distribution $\mathrm{CV}_{\alpha}$.

\subsubsection{Comparison with simulations}

In our theory, we consider disorder-averaged quantities and stationary
processes. To compare the theory with a single simulation, we assume
self-averaging in the sense that the activity distribution across
neurons is approximately the same for each network realization. Since
neurons with different indegrees have different disorder- and time-averaged
inputs, in practice this means that we assume that neurons with comparable
indegree have comparable activity statistics in each network realization.

The disorder averages preserve the static variability across neurons,
as we consider the same connectivity statistics, and in particular
the same indegree distribution, across realizations. Self-averaging
works well when each neuron (or at least a sufficiently large proportion
of neurons) receives input from a representative sample of the rest
of the network.

Under stationarity, distributions across neurons of instantaneous
rates at any given time point (but not of instantaneous rates across
time points --- which we do not consider here) equal distributions
of time-averaged rates across neurons. To obtain the rate distributions
from the simulations, we use time-averaged rates to reduce the variance
of the corresponding estimates. Similarly, we use time averages to
compute the single-neuron autocorrelation functions and power spectra.

We focus on the second-order statistics. Since first-order statistics,
i.e.~the firing rate, scale the power spectra and correlation function
\citep{Gerstner14}, we plot $S_{x}(f)/\nu$ and $C_{x}(f)/\nu^{2}$
to eliminate this trivial dependency. Note that a multiplicative factor
does not influence the intrinsic timescale, \prettyref{eq:def_taucorr}.

\section{Generalized Linear Model Neurons}

First, we consider generalized linear model (GLM) neurons \citep{Gerstner14,Toyoizumi09}.
GLM neurons are stochastic model neurons that spike according to an
inhomogeneous Poisson process at a rate determined by the synaptic
input. Due to their simplicity, GLM neurons are frequently fitted
to experimental data \citep{Gerstner14}; here we consider them because
they are analytically tractable.

\subsection{Neuron dynamics}

Each neuron generates a spike train according to an inhomogeneous
Poisson process with intensity (rate)

\begin{align}
\lambda_{i}^{\alpha}(t) & =c_{1}^{\alpha}\,\phi\left(c_{2}^{\alpha}\left(V_{i}^{\alpha}(t)-\theta^{\alpha}\right)\right)\label{eq:GLM_phi}
\end{align}
where $\theta^{\alpha}$ denotes the (soft) threshold, $\phi(V)$
is a smooth, nonnegative, monotonically increasing function, and $\mbox{\ensuremath{c_{1}^{\alpha}>0}}$,
$\mbox{\ensuremath{c_{2}^{\alpha}>0}}$ are free parameters. The voltage
is given by a linear filtering of the input
\begin{align}
V_{i}^{\alpha}(t) & =\int_{-\infty}^{\infty}\kappa^{\alpha}(t-s)\eta_{i}^{\alpha}\left(s-d^{\alpha\beta}\right)ds\label{eq:GLM_voltage}
\end{align}
where $d^{\alpha\beta}$ allows for a transmission delay. For all
simulations, we choose a filter with a single exponential with time
constant $\tau_{\mathrm{{m}}}^{\alpha}$ which corresponds to post-synaptic
currents in the form of delta spikes:
\begin{align}
\kappa^{\alpha}(t) & =\Theta(t)e^{-t/\tau_{\mathrm{{m}}}^{\alpha}}.\label{eq:GLM_kappa}
\end{align}
Here, $\Theta(t)$ denotes the Heaviside function ensuring causality
of the filter. We rescale the synaptic weights $J_{ij}^{\alpha\beta}$
and the threshold $\theta^{\alpha}$ using $c_{2}^{\alpha}$ such
that $c_{2}^{\alpha}=1$ throughout the rest of this section.

\subsubsection{Colored noise problem}

The effective stochastic input $\eta^{\alpha}(t)$ leads to stochastic
voltage dynamics. Because the voltage is given by a convolution, the
voltage becomes a Gaussian process with
\begin{align}
\mu_{V}^{\alpha}=\bar{{\kappa}}^{\alpha}\mu_{\eta}^{\alpha},\qquad & C_{V}^{\alpha}(\tau)=\int_{-\infty}^{\infty}\tilde{{\kappa}}^{\alpha}(\tau-s)C_{\eta}^{\alpha}(s)ds,\label{eqs:GLM_input_to_voltage}
\end{align}
where the filter determines $\bar{{\kappa}}^{\alpha}=\int_{-\infty}^{\infty}\kappa^{\alpha}(t)dt$
and $\mbox{\ensuremath{\tilde{{\kappa}}^{\alpha}(t)=\int_{-\infty}^{\infty}\kappa^{\alpha}\left(s\right)\kappa^{\alpha}\left(s-t\right)ds}}$.
For the single-exponential filter that we used in simulations, we
have $\bar{{\kappa}}^{\alpha}=\tau_{\mathrm{{m}}}^{\alpha}$ and $\mbox{\ensuremath{\tilde{{\kappa}}^{\alpha}(t)=\frac{\tau_{\mathrm{{m}}}^{\alpha}}{2}e^{-|t|/\tau_{\mathrm{{m}}}^{\alpha}}}}$.
Note that the transmission delay cancels in the stationary case considered
here.

All cumulants of the resulting spike trains $x(t)$ can be obtained
from their characteristic functional \citep{Stratonovich67} (see
Appendix A, \prettyref{eq:appendix_charfctl_Poisson}):
\begin{align*}
\Phi_{x}[u(t)] & =\exp\left(\int_{0}^{T}\left(e^{iu(t)}-1\right)\lambda(t)dt\right).
\end{align*}
From here, we temporarily drop the population index for the sake of
clarity. Averaging over realizations of the rates yields
\begin{align*}
\langle\Phi_{x}[u(t)]\rangle_{\lambda} & \approx e^{\int_{0}^{T}\left(e^{iu(t)}-1\right)\mu_{\lambda}(t)dt}\\
 & \times e^{\frac{1}{2}\int_{0}^{T}\int_{0}^{T}\left(e^{iu(t)}-1\right)C_{\lambda}(t,t^{\prime})\left(e^{iu(t^{\prime})}-1\right)dtdt^{\prime}}
\end{align*}
where we neglect terms of $O(u^{3})$ since we are only interested
in the first and second cumulants. Expanding also $e^{iu(t)}-1$ to
second order in $u(t)$, we can simply read off the stationary cumulants
\begin{align}
\nu & =\mu_{\lambda},\qquad C_{x}(\tau)=\mu_{\lambda}\delta(\tau)+C_{\lambda}(\tau),\label{eqs:GLM_rate_to_spikes}
\end{align}
in agreement with the result of \citep{Krumin09}.

We are left with the task of calculating the first two cumulants of
$\lambda(t)$ from $\mu_{V}$ and $C_{V}(\tau)$, depending on the
choice of the nonlinearity $\phi(V)$.

\subsubsection{Exponential nonlinearity}

First, we consider the commonly employed exponential nonlinearity
\citep{Gerstner14}
\begin{align}
\phi(V) & =\exp\left(V\right).\label{eq:GLM_exp}
\end{align}
Both cumulants are straightforward to obtain from the characteristic
functional of the voltage. We have (see Appendix A, \prettyref{eq:appendix_charfctl_gauss_1}
and \prettyref{eq:appendix_charfctl_gauss_2})
\begin{align*}
\langle\phi(V(t_{1}))\rangle_{V} & =\langle e^{\int_{0}^{T}V(t)\delta(t-t_{1})dt}\rangle_{V}\\
 & =e^{\mu_{V}+\tfrac{1}{2}C_{V}(0)},\\
\langle\phi(V(t_{1}))\phi(V(t_{2}))\rangle_{V} & =\langle e^{\int_{0}^{T}V(t)[\delta(t-t_{1})+\delta(t-t_{2})]dt}\rangle_{V}\\
 & =e^{2\mu_{V}+C_{V}(0)+C_{V}(t_{2}-t_{1})},
\end{align*}
where we used the stationarity of $V$. Including the prefactor and
the threshold from \prettyref{eq:GLM_phi}, we get
\begin{align}
\mu_{\lambda} & =c_{1}\,\exp\left(\mu_{V}-\theta+\tfrac{1}{2}C_{V}(0)\right),\label{eq:GLM_exp_mean}\\
C_{\lambda}(\tau) & =\mu_{\lambda}^{2}\exp(C_{V}(\tau))-\mu_{\lambda}^{2}.\label{eq:GLM_exp_corr}
\end{align}
From \prettyref{eq:GLM_exp_corr} it follows that $C_{\lambda}(\tau)$
has a static part as long as $C_{V}(\infty)>0$. Since $C_{\eta}(\tau)$
contains a static part (see \prettyref{eq:DMFT_static} and \prettyref{eq:DMFT_correlation_pop}),
$C_{V}(\tau)$ and hence $C_{\lambda}(\tau)$ and $C_{x}(\tau)$ indeed
also contain a static contribution and saturate on a plateau.

\paragraph{Rate distribution}

The rate distribution across neurons is lognormal because the (static)
input distribution is Gaussian and the f-I curve is a simple exponential
\citep{Roxin11_16217}. The theory yields the mean $\nu=c_{1}\,\exp\left(\mu_{V}-\theta+\tfrac{1}{2}C_{V}(0)\right)$
and variance $\sigma_{\nu}^{2}=C_{x}(\infty)=\nu^{2}\left(e^{C_{V}(\infty)}-1\right)$
of the firing rate. We note that we can obtain the same result from
a constant input with mean $\tilde{\mu}_{V}=\mu_{V}-\theta+\tfrac{1}{2}C_{V}(0)-\frac{1}{2}C_{V}(\infty)$
and variance across neurons $\tilde{\sigma}_{V}^{2}=C_{V}(\infty)$.
Parameterized in terms of $\tilde{\mu}_{V}$ and $\tilde{\sigma}_{V}$,
the firing rate distribution is thus
\begin{align}
p(\nu) & =\nu^{-1}\,\mathcal{N}(\ln(\nu/c_{1})\,|\,\tilde{\mu}_{V},\tilde{\sigma}_{V}^{2})\label{eq:GLM_exp_ratedist}
\end{align}
with the normal distribution $\mathcal{N}(x\,|\,\mu,\sigma^{2})$.

\subsubsection{Error function nonlinearity}

A drawback of the exponential function, \prettyref{eq:GLM_exp}, is
that it allows for infinite rates. Thus, we also consider the bounded
nonlinearity
\begin{align}
\phi(V) & =\frac{1}{2}\left(1+\erf\left(V/\sqrt{2}\right)\right).\label{eq:GLM_erf}
\end{align}
The integrals to determine the cumulants can be solved using the table
\citep{Owen80_389} (details in Appendix B.1); the result is
\begin{align}
\mu_{\lambda} & =\frac{c_{1}}{2}\left(1+\erf\left(h/\sqrt{2}\right)\right),\label{eq:GLM_erf_mean}\\
C_{\lambda}(\tau) & =c_{1}\mu_{\lambda}-2c_{1}^{2}T\left(h,a(\tau)\right)-\mu_{\lambda}^{2},\label{eq:GLM_erf_corr}
\end{align}
where we again suppressed the population index, abbreviated $h=\frac{\mu_{V}-\theta}{\sqrt{1+C_{V}(0)}}$
and $a(\tau)=\left(\frac{1+C_{V}(0)-C_{V}(\tau)}{1+C_{V}(0)+C_{V}(\tau)}\right)^{1/2}$,
and used Owen's T function $T(h,a)=\frac{1}{2\pi}\int_{0}^{a}dx\,\frac{e^{-\frac{1}{2}h^{2}(1+x^{2})}}{1+x^{2}}$.

\paragraph{Rate distribution}

Equivalent to the situation for the exponential nonlinearity, the
input distribution across neurons is Gaussian. Again, we consider
the equivalent static problem which, in this case, leads to $\mbox{\ensuremath{\tilde{\mu}_{V}=\frac{\mu_{V}-\theta}{\sqrt{1+C_{V}(0)-C_{V}(\infty)}}}}$
and $\tilde{\sigma}_{V}^{2}=\frac{C_{V}(\infty)}{1+C_{V}(0)-C_{V}(\infty)}$.
Parameterized in terms of $\tilde{\mu}_{V}$ and $\tilde{\sigma}_{V}$,
the firing rate distribution is
\begin{align}
p(\nu) & =\frac{\mathcal{N}(\probit(\nu/c_{1})\,|\,\tilde{\mu}_{V},\tilde{\sigma}_{V}^{2})}{c_{1}\,\mathcal{N}(\probit(\nu/c_{1})\,|\,0,1)},\label{eq:GLM_erf_ratedist}
\end{align}
where $\probit(x)$ denotes the inverse of the standard normal cumulative
distribution, i.e.~$\probit(\phi(V))=V$, and we used $\phi^{\prime}(V)=\mathcal{N}(V\,|\,0,1)$.

\subsubsection{Numerical solution of the self-consistency problem}

We solve the self-consistency problem using a fixed-point iteration
\citep{Lerchner06_131,dummer14}. To initiate the algorithm, we set
$\nu^{\alpha}=\frac{1}{2}c_{1}^{\alpha}$ and $C_{\lambda}^{\alpha}(t)=0$.
Next, we determine the input statistics according to \prettyref{eq:DMFT_mean_pop}
and \prettyref{eq:DMFT_correlation_pop}; then we determine the voltage
statistics according to \prettyref{eqs:GLM_input_to_voltage}. From
the voltage statistics we can obtain the statistics of the rate via
\prettyref{eq:GLM_exp_mean} and \prettyref{eq:GLM_exp_corr} (or
\prettyref{eq:GLM_erf_mean} and \prettyref{eq:GLM_erf_corr}). Denoting
the rate thus calculated as $\hat{\mu}_{\lambda,n+1}^{\alpha}$, we
then update the rate statistics using incremental steps, $\mu_{\lambda,n+1}^{\alpha}=\mu_{\lambda,n}^{\alpha}+\varepsilon(\hat{\mu}_{\lambda,n+1}^{\alpha}-\mu_{\lambda,n}^{\alpha})$
for the mean rate, and similarly for all entries of $C_{\lambda}^{\alpha}(t)$.
The new firing rate statistics lead via \prettyref{eqs:GLM_rate_to_spikes}
to new spike train statistics. Here, the small update step $\varepsilon<1$
is crucial because otherwise the fixed-point iteration is numerically
unstable. Now we iterate and generate new voltage statistics. With
the incremental update and the initialization $\nu^{\alpha}=\frac{1}{2}c_{1}^{\alpha}$,
the algorithm quickly converged to the fixed point corresponding to
the simulation in the examples we considered. Due to the analytical
solutions, the only bottleneck for the numerics is the convolution
in \prettyref{eqs:GLM_input_to_voltage}, which can be solved efficiently
using the Fast Fourier Transform \citep{Press07}. Thus, even the
parameter scans with $5,\!000$ points described in the following
run on a laptop in less than two minutes.

\subsection{Balanced random network}

As a first application of the theory, we consider a balanced random
network of excitatory cells and inhibitory interneurons. The network
contains two populations (\prettyref{fig:BRN-GLM-EXP}\textbf{a}),
$\alpha\in\{E,I\}$, and it is driven by an excitatory external input
which we incorporate into an effective threshold $\theta_{\mathrm{eff}}=\theta-\mu_{\mathrm{ext}}$.
Here, we use a constant external input rather than a Poisson drive
because we are particularly interested in finding long timescales,
which might be hindered by the lack of temporal correlation of Poisson
spike trains. However, the theory can straightforwardly be applied
to Poisson input. Although four times more excitatory cells are present
in the network, we typically place it in an inhibition-dominated regime
by increasing the synaptic weights of the inhibitory neurons. As well
known \citep{Brunel00}, this settles the network in the balanced
state leading to asynchronous irregular activity of the neurons (see,
e.g.,~\prettyref{fig:BRN-GLM-EXP}\textbf{b}).

In line with Brunel's model A \citep{Brunel00}, we choose identical
values for the single-neuron parameters. Since we also choose the
same connection probability of $10\%$ for all pairs of populations,
both populations receive statistically identical input in the DMFT
approximation. Due to identical single-neuron parameters and input
statistics, the statistics of the activity is the same for excitatory
and inhibitory neurons (see, e.g.,~\prettyref{fig:BRN-GLM-EXP}\textbf{b});
therefore, we do not distinguish between the populations for the statistics
in our plots. In contrast to the network examined by Brunel, we consider
the somewhat more involved case of a fixed connection probability
between a pair of neurons instead of a fixed number of incoming synapses
per neuron (indegree). The fixed connection probability leads to a
(binomially) distributed indegree across neurons, such that a strong
variability across neurons is present in the network (see, e.g.,~\prettyref{fig:BRN-GLM-EXP}\textbf{c}).
This variability is already present on the level of mean firing rates,
i.e.~there is static variability in the network.

All simulations were performed using the NEST simulator version 2.20.1
\citep{Nest2201}. In all GLM network simulations, we simulated $1\:\mathrm{min}$
of biological time with a time step of $0.1\:\mathrm{ms}$ and discarded
an initial transient of $1\:\mathrm{s}$. For the GLM neurons, we
used the `pp\_psc\_delta' neuron model. To allow for the error function
nonlinearity, we modified the `pp\_psc\_delta' model accordingly.

\subsubsection{Exponential nonlinearity: absence of long timescales}

\begin{figure}
\centering{}\includegraphics[width=3.375in]{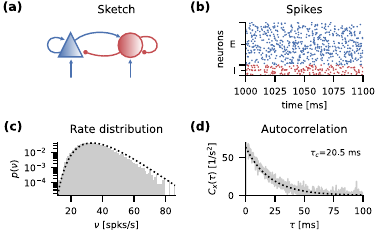}\caption{Balanced random network of GLM neurons with exponential nonlinearity.
\textbf{(a)} Sketch of the network with populations of excitatory
(blue) and inhibitory (red) neurons. \textbf{(b)} Raster plot of 2\%
of the excitatory (blue) and inhibitory (red) neurons. \textbf{(c)}
Firing rate distribution across all neurons from simulation (gray)
and theory (black) using \prettyref{eq:GLM_exp_ratedist}. \textbf{(d)}
Population-averaged single-unit autocorrelation function from simulation
(gray) and self--consistent theory (black) using \prettyref{eq:GLM_exp_mean}
and \prettyref{eq:GLM_exp_corr}. Here, we subtracted the static contribution
$C_{x}(\infty)$. Parameters: $N_{E}=10,\!000$, $N_{I}=2,\!500$,
$J_{E}=0.25\:\mathrm{mV}$, $|J_{I}/J_{E}|=4.5$, $p=0.1$, $\tau_{\mathrm{m}}=20\:\mathrm{ms}$,
$\theta_{\mathrm{eff}}=0\:\mathrm{mV}$, $c_{1}=50\:\mathrm{s}^{-1}$,
$c_{2}=0.02\:\text{\ensuremath{\mathrm{mV}^{-1}}}$, $d=1.5\:\mathrm{ms}$.
\label{fig:BRN-GLM-EXP}}
\end{figure}
First, we consider networks with an exponential nonlinearity (\prettyref{fig:BRN-GLM-EXP}).
The fixed-point iteration yields a rate distribution and an autocorrelation
function that closely match the simulation (\prettyref{fig:BRN-GLM-EXP}\textbf{c},\textbf{d}).
The theory for the rate distribution (\prettyref{fig:BRN-GLM-EXP}\textbf{c})
is slightly biased towards higher rates; a possible cause for this
is a finite size effect because the mean inhibitory indegree $K_{I}=pN_{I}=250$
is relatively small. Nonetheless, the theory predicts the autocorrelation
function very well (\prettyref{fig:BRN-GLM-EXP}\textbf{d}) and yields
a timescale $\tau_{c}\approx\tau_{\mathrm{{m}}}=20\,\mathrm{ms}$.

\begin{figure}
\includegraphics{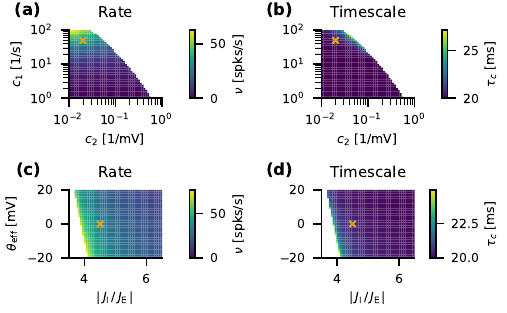}

\caption{Parameter scan for a balanced random network of GLM neurons with exponential
nonlinearity. \textbf{(a},\textbf{b)} Firing rate and intrinsic timescale
for varying neuron parameters $c_{1}$ and $c_{2}$. Parameters used
in (\textbf{c},\textbf{d}) and \prettyref{fig:BRN-GLM-EXP} indicated
by orange crosses. \textbf{(c},\textbf{d)} Firing rate and intrinsic
timescale for varying effective threshold $\theta_{\mathrm{eff}}$
and relative inhibitory strength $|J_{I}/J_{E}|$. Parameters used
in (\textbf{a},\textbf{b}) and \prettyref{fig:BRN-GLM-EXP} indicated
by orange crosses. Further parameters as in \prettyref{fig:BRN-GLM-EXP}.
\label{fig:GLM-EXP-SCAN}}
\end{figure}

For the parameters in \prettyref{fig:BRN-GLM-EXP}, the intrinsic
timescale is close to the membrane time constant. This raises the
question whether longer timescales can be achieved in a network of
GLM neurons. To answer this question, we employ our theory and perform
parameter scans. First, we vary the single-neuron parameters $c_{1}$
and $c_{2}$ (\prettyref{fig:GLM-EXP-SCAN}\textbf{a},\textbf{b}).
The rate increases monotonically with $c_{1}$ while $c_{2}$ has
as smaller effect up to a certain threshold (\prettyref{fig:GLM-EXP-SCAN}\textbf{a}).
Beyond this threshold, the rate diverges rapidly to infinity in the
threshold iteration (white area in \prettyref{fig:GLM-EXP-SCAN}\textbf{a}).
The timescale is close to the membrane time constant throughout the
non-divergent regime and only increases slightly towards the threshold
where the rate diverges (\prettyref{fig:GLM-EXP-SCAN}\textbf{b}).
Next, we vary the strength of the external input by adjusting the
effective threshold $\theta_{\mathrm{eff}}$ and the inhibition dominance
by varying $|J_{I}/J_{E}|$ for constant $J_{E}$. We find a clear
threshold of $|J_{I}/J_{E}|$ beyond which the rate diverges (\prettyref{fig:GLM-EXP-SCAN}\textbf{c}).
Again, this threshold corresponds to the regime where the timescale
slowly starts to grow above the membrane time constant.

Put together, these observations suggest that the rate divergence
prevents recurrent dynamics with long timescales in balanced random
networks of GLM neurons with exponential nonlinearity.

\subsubsection{Error function nonlinearity: existence of long timescales}

\begin{figure}
\includegraphics{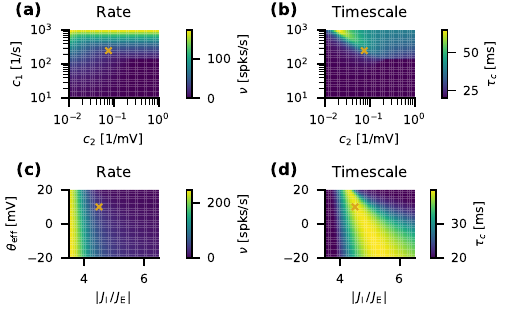}

\caption{Parameter scan for a balanced random network of GLM neurons with error
function nonlinearity. \textbf{(a},\textbf{b)} Firing rate and intrinsic
timescale for varying neuron parameters $c_{1}$ and $c_{2}$. Parameters
used in (\textbf{c},\textbf{d}) and \prettyref{fig:BRN-GLM-ERF} indicated
by orange crosses. \textbf{(c},\textbf{d)} Firing rate and intrinsic
timescale for varying effective threshold $\theta_{\mathrm{eff}}$
and relative inhibitory strength $|J_{I}/J_{E}|$. Parameters used
in (\textbf{a},\textbf{b}) and \prettyref{fig:BRN-GLM-ERF} indicated
by orange crosses. Further parameters as in \prettyref{fig:BRN-GLM-EXP}.
\label{fig:GLM-ERF-SCAN}}
\end{figure}

In the previous section, the rate divergence prevented long timescales.
To avoid the divergence, we consider the bounded transfer function
\prettyref{eq:GLM_erf} and use our theory for parameter scans (\prettyref{fig:GLM-ERF-SCAN}).
The effect of the single-neuron parameters $c_{1}$ and $c_{2}$ is
similar to the unbounded case but the rate divergence is absent (\prettyref{fig:GLM-ERF-SCAN}\textbf{a}).
This allows for a parameter regime with longer timescales up to approximately
$3\tau_{\mathrm{{m}}}$ (\prettyref{fig:GLM-ERF-SCAN}\textbf{b}).
Similarly, varying $\theta_{\mathrm{eff}}$ and $|J_{I}/J_{E}|$ uncovers
a regime with a rate close to the maximum $c_{1}$ when the network
is not inhibition-dominated (\prettyref{fig:GLM-ERF-SCAN}\textbf{c}).
Outside the inhibition-dominated regime, we expect that our theory
does not yield quantitatively accurate predictions. The effect on
the timescale is more subtle: within the inhibition-dominated regime,
for any given $|J_{I}/J_{E}|$ the timescale displays a maximum whose
location depends on the external input (\prettyref{fig:GLM-ERF-SCAN}\textbf{d}).

\begin{figure}
\includegraphics{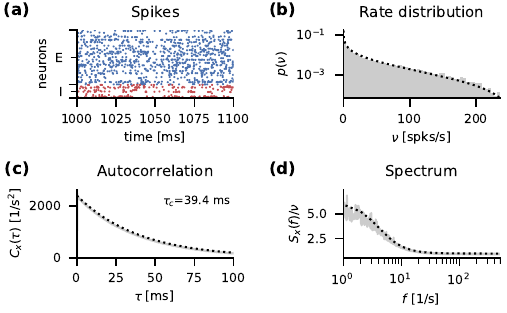}

\caption{Balanced random network of GLM neurons with error function nonlinearity.
\textbf{(a)} Raster plot of 2\% of the excitatory (blue) and inhibitory
(red) neurons. \textbf{(b)} Firing rate distributions across all neurons
from simulation (gray) and theory (black) using \prettyref{eq:GLM_erf_ratedist}.
\textbf{(c},\textbf{d)} Population-averaged single-unit autocorrelation
function and power spectrum from simulation (gray) and self--consistent
theory (black) using \prettyref{eq:GLM_erf_mean} and \prettyref{eq:GLM_erf_corr}.
As in \prettyref{fig:BRN-GLM-EXP}, we subtracted the static contribution
$C_{x}(\infty)$. Parameters: $c_{1}=250\:\mathrm{s}^{-1}$, $c_{2}=0.075\:\text{\ensuremath{\mathrm{mV}^{-1}}}$,
further parameters as in \prettyref{fig:BRN-GLM-EXP}. \label{fig:BRN-GLM-ERF}}
\end{figure}

What kind of dynamics is displayed by the network at such a local
maximum of the timescale? The corresponding spike trains show a strong
variability of firing rate across neurons and temporally correlated
spikes (\prettyref{fig:BRN-GLM-ERF}\textbf{a}). The rate distribution
reveals that all rates between the minimum zero and the maximum $c_{1}$
are present, in excellent agreement with the theoretical prediction
(\prettyref{fig:BRN-GLM-ERF}\textbf{b}). In the example considered,
the empirical estimate of the network--averaged single-unit autocorrelation
displays an intrinsic timescale of approximately $2\tau_{\mathrm{{m}}}$;
again, the empirical estimate and the theoretical prediction agree
closely (\prettyref{fig:BRN-GLM-ERF}\textbf{c}). From the spike train
power spectrum, a high $\mathrm{CV}>2$ is apparent (\prettyref{fig:BRN-GLM-ERF}\textbf{d}).
All of these characteristics agree with the ``heterogeneous asynchronous
state'' uncovered in \citep{Ostojic14}.

\subsubsection{Error function nonlinearity: mechanism of timescale}

To uncover the mechanisms that shape the timescale, in particular
the local maximum in \prettyref{fig:GLM-ERF-SCAN}\textbf{d}, we develop
a theory for the asymptotic timescale $\tau_{\infty}$, \prettyref{eq:def_tauinf}.
To this end, we use that $\tilde{{\kappa}}(t)=\frac{\tau_{\mathrm{{m}}}}{2}e^{-|t|/\tau_{\mathrm{{m}}}}$
is the fundamental solution to the differential operator $1-\tau_{\mathrm{{m}}}^{2}\frac{d^{2}}{dt^{2}}$,
i.e.~$\left(1-\tau_{\mathrm{{m}}}^{2}\frac{d^{2}}{dt^{2}}\right)\tilde{{\kappa}}(t)=\tau_{\mathrm{{m}}}^{2}\delta(t)$.
Thus, we can rewrite \prettyref{eqs:GLM_input_to_voltage} into a
differential equation:
\begin{align*}
\tau_{\mathrm{{m}}}^{2}\ddot{C}_{V} & =C_{V}-\tau_{\mathrm{{m}}}^{2}C_{\eta}
\end{align*}
where the dependence of $C_{\eta}$ on $C_{V}$ is determined by \prettyref{eq:DMFT_correlation_pop},
\prettyref{eqs:GLM_rate_to_spikes}, and \prettyref{eq:GLM_erf_corr}.
Next, we rescale time such that $\tau_{\mathrm{{m}}}=1$ and linearize
this differential equation for small $\Delta_{V}(\tau)\equiv C_{V}(\tau)-C_{V}(\infty)$
to obtain
\begin{align*}
\ddot{\Delta}_{V} & =\left(1-\frac{dC_{\eta}(\infty)}{dC_{V}(\infty)}\right)\Delta_{V}+O(\Delta_{V}^{2}).
\end{align*}
This allows for an exponential solution with time constant
\begin{equation}
\tau_{\infty}=\frac{1}{\sqrt{1-g^{2}\frac{dC_{\lambda}(\infty)}{dC_{V}(\infty)}}}\label{eq:GLM_tau_inf}
\end{equation}
where we used \prettyref{eq:DMFT_correlation_pop} and \prettyref{eqs:GLM_rate_to_spikes}
to derive $\mbox{\ensuremath{\frac{dC_{\eta}(\infty)}{dC_{V}(\infty)}=g^{2}\frac{dC_{\lambda}(\infty)}{dC_{V}(\infty)}}}$.
We see that there are two factors that determine the timescale: the
cumulant of the connectivity $g^{2}$ and the gain of the rate autocorrelation
$\frac{dC_{\lambda}(\infty)}{dC_{V}(\infty)}$. For the latter, we
obtain from \prettyref{eq:GLM_erf_corr}
\begin{align}
\frac{dC_{\lambda}(\infty)}{dC_{V}(\infty)} & =\frac{c_{1}^{2}}{2\pi}\frac{\exp\left(-\frac{(\mu_{V}-\theta_{\mathrm{eff}})^{2}}{1+C_{V}(0)+C_{V}(\infty)}\right)}{\sqrt{(1+C_{V}(0))^{2}-C_{V}(\infty)^{2}}}.\label{eq:GLM_erf_gain}
\end{align}
Thus, given a self--consistent autocorrelation $C_{x}$ and the corresponding
voltage statistics from \prettyref{eqs:GLM_input_to_voltage}, the
asymptotic timescale \prettyref{eq:GLM_tau_inf} can be directly evaluated.

\begin{figure}
\includegraphics{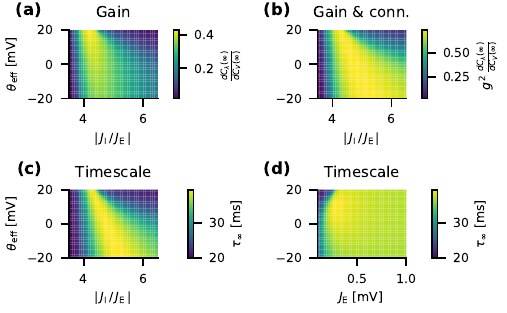}

\caption{Mechanisms that shape the asymptotic timescale. \textbf{(a)} Asymptotic
gain $\frac{dC_{\lambda}(\infty)}{dC_{V}(\infty)}$ of the rate autocorrelation
w.r.t.~changes in the voltage autocorrelation, \prettyref{eq:GLM_erf_gain}.
\textbf{(b)} Asymptotic gain multiplied by the second cumulant of
the connectivity, $g^{2}$. \textbf{(c)} Asymptotic timescale according
to \prettyref{eq:GLM_tau_inf}, $\tau_{\infty}=\left(1-g^{2}\frac{dC_{\lambda}(\infty)}{dC_{V}(\infty)}\right)^{-1/2}$,
for varying effective threshold $\theta_{\mathrm{eff}}$ and relative
inhibitory strength $|J_{I}/J_{E}|$. \textbf{(d)} Same as \textbf{c}
for varying excitatory synaptic strength $J_{E}$ with constant $|J_{I}/J_{E}|$.
Further parameters as in \prettyref{fig:BRN-GLM-ERF}. \label{fig:GLM_ERF_mechanism}}
\end{figure}

We vary $\theta_{\mathrm{eff}}$ and $|J_{I}/J_{E}|$ in \prettyref{fig:GLM_ERF_mechanism}\textbf{a}-\textbf{c}.
First, we plot $\frac{dC_{\lambda}(\infty)}{dC_{V}(\infty)}$ alone,
which we refer to as the gain (\prettyref{fig:GLM_ERF_mechanism}\textbf{a}).
Due to the interplay between the exponential suppression $\exp\left(-\frac{(\mu_{V}-\theta_{\mathrm{eff}})^{2}}{1+C_{V}(0)+C_{V}(\infty)}\right)$
and the square root factor $1/\sqrt{(1+C_{V}(0))^{2}-C_{V}(\infty)^{2}}<1$
in \prettyref{eq:GLM_erf_gain}, the gain already exhibits a maximum.
The existence of the maximum is mainly determined by the exponential
suppression with growing $|\mu_{V}-\theta_{\mathrm{eff}}|$ in \prettyref{eq:GLM_erf_gain}:
both in the excitation- and in the inhibition-dominated regime, $\mu_{V}$
is far from the effective threshold $\theta_{\mathrm{eff}}$. The
precise location of the maximum is not necessarily at $\mu_{V}=\theta_{\mathrm{eff}}$
as it is also determined by the square root factor. The latter decays
reciprocally to $C_{V}(0)$ and $C_{V}(\infty)$. Both $C_{V}(0)$
and $C_{V}(\infty)$ decay for growing effective threshold and inhibition
dominance, which results in a larger square root factor that shifts
the maximum towards the upper right and broadens it. The cumulant
of the connectivity $g^{2}$ grows with $|J_{I}/J_{E}|^{2}$, which
further broadens the region of the maximum (\prettyref{fig:GLM_ERF_mechanism}\textbf{b}).
The resulting asymptotic timescale (\prettyref{fig:GLM_ERF_mechanism}\textbf{c})
agrees both qualitatively and quantitatively with the intrinsic timescale
(\prettyref{fig:GLM-ERF-SCAN}\textbf{d}). This is likely due to the
single-exponential shape of the autocorrelation function (\prettyref{fig:BRN-GLM-ERF}\textbf{c}).

To investigate the interplay of the gain and the connectivity further,
we vary the overall synaptic strengths by varying the excitatory weight
$J_{E}$ while keeping $|J_{I}/J_{E}|$ fixed at an inhibition-dominated
value (\prettyref{fig:GLM_ERF_mechanism}\textbf{d}). Increasing $J_{E}$
in the inhibition-dominated regime shifts $\mu_{V}$ away from the
effective threshold and decreases the gain; conversely $g^{2}$ grows
with $J_{E}^{2}$. This interplay leads to a broad region in parameter
space with an increased timescale. However, the exponential decrease
of the gain is more pronounced than the quadratic increase of $g^{2}$
such that the asymptotic timescale does not continue to grow with
$J_{E}$ but saturates. Thus, although increasing $J_{E}$ goes together
with increased variability across neurons as in the ``heterogeneous
asynchronous state'' described by Ostojic \citep{Ostojic14}, this
does not map systematically onto longer single-neuron timescales.

\subsubsection{Error function nonlinearity: external timescale}

Our theory allows arbitrary Gaussian processes as external input.
To investigate the influence of an external timescale on the intrinsic
timescale, we choose a zero-mean Ornstein-Uhlenbeck process with
\begin{equation}
C_{\mathrm{ext}}(\tau)=\frac{\sigma_{\mathrm{ext}}^{2}}{\tau_{\mathrm{{m}}}}\left(\frac{1}{\tau_{\mathrm{{m}}}}+\frac{1}{\tau_{\mathrm{ext}}}\right)e^{-|\tau|/\tau_{\mathrm{ext}}}.\label{eq:GLM_ext_corrfct}
\end{equation}
Here, the scaling factors ensure that the external timescale does
not influence the resulting variance of the voltage, $C_{V}(0)=\int_{-\infty}^{\infty}\tilde{{\kappa}}(s)C_{\mathrm{ext}}(s)ds=\sigma_{\mathrm{ext}}^{2}$
for $\mbox{\ensuremath{\tilde{{\kappa}}(t)=\frac{1}{2}\tau_{\mathrm{{m}}}e^{-|t|/\tau_{\mathrm{{m}}}}}}$.

\begin{figure}
\includegraphics{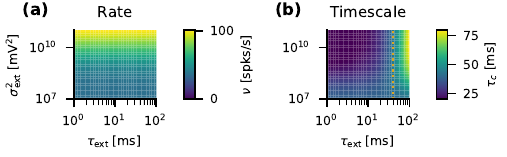}

\caption{Influence of colored external input. \textbf{(a},\textbf{b)} Firing
rate and intrinsic timescale for varying strength $\sigma_{\mathrm{ext}}^{2}$
and timescale $\tau_{\mathrm{ext}}$ of an external Ornstein-Uhlenbeck
process. Orange line in \textbf{(b)} indicates the intrinsic timescale
without external input. Parameters as in \prettyref{fig:BRN-GLM-ERF}.
\label{fig:GLM_ERF_tauext}}
\end{figure}

We take the parameters from \prettyref{fig:BRN-GLM-ERF} where the
intrinsic timescale is maximal in the absence of external input. Increasing
the strength of the external input $\sigma_{\mathrm{ext}}^{2}$ leads
to an increased firing rate (\prettyref{fig:GLM_ERF_tauext}\textbf{a}).
As desired, by construction of \prettyref{eq:GLM_ext_corrfct}, the
external timescale has a negligible effect on the firing rate at constant
$\sigma_{\mathrm{ext}}^{2}$ (\prettyref{fig:GLM_ERF_tauext}\textbf{a}).
The effect of the external timescale on the intrinsic timescale is
highly intuitive: If $\tau_{\mathrm{ext}}$ is smaller than the intrinsic
timescale without external input it decreases the intrinsic timescale,
and vice versa (\prettyref{fig:GLM_ERF_tauext}\textbf{b}). The strength
of this effect grows with the strength of the external input. In the
limit of strong external input, the intrinsic timescale approaches
the external timescale if $\tau_{\mathrm{ext}}>\tau_{\mathrm{{m}}}$;
if $\tau_{\mathrm{ext}}<\tau_{\mathrm{{m}}}$ the intrinsic timescale
approaches the minimum set by the membrane time constant.

\section{Leaky Integrate-and-Fire Neurons}

Considering GLM neurons is a convenient choice due to their analytical
tractability. However, their intrinsic stochasticity might fundamentally
alter the network dynamics. Thus, we consider the frequently used
leaky integrate--and--fire neuron model in this section \citep{Gerstner14}.
The synapses are taken to be current-based with an exponential time
course. An analytical solution to the colored noise problem for LIF
neurons is an open challenge. Here, we focus on the fluctuation-driven
regime and employ an approach based on the Wiener--Rice series \citep{Rice45,Ricciardi83_454,Verechtchaguina06_031108}
and the Stratonovich approximation thereof \citep{Stratonovich67,Verechtchaguina06_031108}.
Below, we briefly introduce both the Wiener--Rice series and its
Stratonovich approximation. For a comprehensive and pedagogic introduction
to this approach, in particular with a focus on LIF neurons, see \citep{Schwalger_unpublished}.

\subsection{Neuron dynamics}

The dynamics of individual neurons are governed by
\begin{align}
\tau_{\mathrm{{m}}}^{\alpha}\dot{V}_{i}^{\alpha}(t) & =-V_{i}^{\alpha}(t)+I_{i}^{\alpha}(t),\label{eq:LIF_voltage}\\
\tau_{\mathrm{{s}}}^{\alpha}\dot{I}_{i}^{\alpha}(t) & =-I_{i}^{\alpha}(t)+\tau_{\mathrm{{m}}}^{\alpha}\eta_{i}^{\alpha}\left(t-d^{\alpha\beta}\right),\label{eq:LIF_current}
\end{align}
where $V_{i}^{\alpha}$ denotes the membrane voltage, $I_{i}^{\alpha}$
the synaptic current, $\tau_{\mathrm{{m/s}}}^{\alpha}$ the membrane/synaptic
time constant, and the voltage is reset to $V_{\mathrm{{r}}}^{\alpha}$
and held constant during the refractory period $\tau_{\mathrm{{ref}}}^{\alpha}$
whenever it reaches the threshold $\theta^{\alpha}$. Threshold crossing
triggers a spike which arrives at another neuron after a delay $d^{\alpha\beta}$.
We set the resting potential to zero without loss of generality and
absorb the membrane resistance into the synaptic current.

\subsubsection{Effective stochastic dynamics}

The effective stochastic input with statistics governed by \prettyref{eq:DMFT_mean_pop}
and \prettyref{eq:DMFT_correlation_pop} leads to a stochastic current
with
\begin{align}
\mu_{I}^{\alpha} & =\tau_{\mathrm{{m}}}^{\alpha}\mu_{\eta}^{\alpha},\label{eq:LIF_current_mean}\\
C_{I}^{\alpha}(\tau) & =\left(\tfrac{\tau_{\mathrm{{m}}}^{\alpha}}{\tau_{\mathrm{{s}}}^{\alpha}}\right)^{2}\int_{-\infty}^{\infty}\tilde{{\kappa}}^{\alpha}(\tau-s)C_{\eta}^{\alpha}(s)ds,\label{eq:LIF_current_corr}
\end{align}
where $\tilde{{\kappa}}^{\alpha}(t)=\frac{\tau_{\mathrm{{s}}}^{\alpha}}{2}e^{-|t|/\tau_{\mathrm{{s}}}^{\alpha}}$,
similar to \prettyref{eqs:GLM_input_to_voltage}. Contrary to the
GLM neurons, the voltage cannot become a stationary process for LIF
neurons due to the fire--and--reset rule. To circumvent this problem,
we use the Wiener--Rice series which relates the free process without
reset to the spiking statistics.

\subsubsection{Wiener--Rice series and Stratonovich approximation}

We consider a LIF neuron after the refractory period and the voltage
dynamics that results if we do not allow for another fire--and--reset.
We denote this free voltage $U(t)$. Moreover, we temporarily neglect
the static contribution to the input variability and drop the population
index. The process starts at $U(0)=V_{\mathrm{{r}}}$ and produces
a system of random points $\{t_{i}\}$ defined by the upcrossings
$U(t_{i})=\theta$, $\dot{U}(t_{i})>0$. For this system of random
points, the probability that no point falls in the interval $[0,T]$,
i.e.~the survival probability, is given by \citep{Stratonovich67}
\begin{align*}
S(T) & =\exp\left(\sum_{s=1}^{\infty}\frac{(-1)^{s}}{s!}\int_{0}^{T}\dots\int_{0}^{T}g_{s}(t_{1},\dots,t_{s})dt_{1}\dots dt_{s}\right)
\end{align*}
where the $g_{s}(t_{1},\dots,t_{s})$ are related to the free upcrossing
probabilities $n_{s}(t_{1},\dots,t_{s})$ calculated below, similar
to the relation between moments and cumulants. For example, $g_{1}(t_{1})=n_{1}(t_{1})$
and $\mbox{\ensuremath{g_{2}(t_{1},t_{2})=n_{2}(t_{1},t_{2})-n_{1}(t_{1})n_{1}(t_{2})}}$.
Now we approximate the output process as a renewal process such that
the survival probability is sufficient to describe the statistics.
Instead of the survival probability, it is more convenient to consider
the cumulative hazard $\mbox{\ensuremath{H(T)=-\ln S(T)}}$ \citep{Gerstner14},
i.e.
\begin{align*}
H(T) & =\sum_{s=1}^{\infty}\frac{(-1)^{s-1}}{s!}\int_{0}^{T}\dots\int_{0}^{T}g_{s}(t_{1},\dots,t_{s})dt_{1}\dots dt_{s}.
\end{align*}
This can be regarded as a resummation of the Wiener--Rice series
in terms of the $g_{s}(t_{1},\dots,t_{s})$ instead of the free upcrossing
probabilities $n_{s}(t_{1},\dots,t_{s})$ \citep{Verechtchaguina06_031108}.

Calculating the free upcrossing probabilities $n_{s}(t_{1},\dots,t_{s})$,
and thus the $g_{s}(t_{1},\dots,t_{s})$, is tedious. To avoid this
difficulty, Stratonovich proposed the approximation \citep{Stratonovich67}
\begin{align}
H_{S}(T) & =-\int_{0}^{T}n_{1}(t)\frac{\ln\left(1-\int_{0}^{T}Q(t,t^{\prime})n_{1}(t^{\prime})dt^{\prime}\right)}{\int_{0}^{T}Q(t,t^{\prime})n_{1}(t^{\prime})dt^{\prime}}dt\label{eq:LIF_Stratonovich_approximation}
\end{align}
where $Q(t_{1},t_{2})=1-\frac{n_{2}(t_{1},t_{2})}{n_{1}(t_{1})n_{1}(t_{2})}$.
Briefly, to derive this approximation, the $g_{s}(t_{1},\dots,t_{s})$
for $s\ge3$ are expressed in terms of $n_{1}(t_{1})$ and $Q(t_{1},t_{2})$
such that both the symmetry of the time arguments $t_{1},\dots,t_{s}$
and the equal-time limit $g_{s}(t_{1},\dots,t_{1})=(-1)^{s-1}(s-1)!\,n_{1}(t_{1})^{s}$
are fulfilled; the resulting approximated $g_{s}(t_{1},\dots,t_{s})$
are inserted into $H(T)$, which leads to a series that can be evaluated
and yields \prettyref{eq:LIF_Stratonovich_approximation}. The condition
$g_{s}(t_{1},\dots,t_{1})=(-1)^{s-1}(s-1)!\,n_{1}(t_{1})^{s}$ holds
for a system of nonapproaching points where $n_{s}(t_{1},\dots,t_{1})=0$
for $s\ge2$, hence \prettyref{eq:LIF_Stratonovich_approximation}
is an approximation constructed for such a system. Although this seems
intuitively reasonable because the voltage dynamics is continuous
and differentiable, this condition is violated for LIF neurons with
exponential postsynaptic currents \citep{Schwalger_unpublished}.
Nonetheless, it yields good results, as shown in the following.

A much simpler alternative to the Stratonovich approximation would
be to set $g_{s}(t_{1},\dots,t_{s})=0$ for $s\ge2$, leading to $H(T)=\int_{0}^{T}n_{1}(t)dt$.
This approximation is sometimes referred to as the Hertz approximation.
In particular, the Hertz approximation leads to a closed expression
for the hazard function $\mbox{\ensuremath{h(t)\equiv\frac{d}{dt}H(t)=n_{1}(t)}}$.
Unfortunately, this approximation is too severe and strongly affects
the resulting firing rate. The main difference between the two approximations
is the asymptotic saturation of the hazard function. Thus, we employ
an approximation suggested by Stratonovich for long times \citep{Stratonovich67}:
$\int_{0}^{T}Q(t,t^{\prime})n_{1}(t^{\prime})dt^{\prime}\approx n_{0}\int_{0}^{\infty}Q(t,t^{\prime})dt^{\prime}\approx n_{0}\eta$
with $n_{0}=\lim_{t\to\infty}n_{1}(t)$ and $\eta=\lim_{t\to\infty}\int_{0}^{\infty}Q(t,t^{\prime})dt^{\prime}$.
Inserting this approximation into \prettyref{eq:LIF_Stratonovich_approximation}
leads to
\begin{equation}
h_{S}(t)=\frac{\kappa_{S}}{n_{0}}n_{1}(t),\qquad\kappa_{S}=-\frac{1}{\eta}\ln\left(1-n_{0}\eta\right).\label{eq:LIF_Hertz_rescaled}
\end{equation}
\prettyref{eq:LIF_Hertz_rescaled} combines the simplicity of the
Hertz approximation with the asymptotic behavior of the Stratonovich
approximation. The asymptotic level is given by $\lim_{t\to\infty}h_{S}(t)=\kappa_{S}$;
to leading order in $\eta$ we have $\mbox{\ensuremath{\kappa_{S}=n_{0}+O(\eta)}}$,
which recovers the Hertz approximation. In the parameter regime we
consider, \prettyref{eq:LIF_Hertz_rescaled} yields very similar results
to \prettyref{eq:LIF_Stratonovich_approximation} (see Appendix D).
In all figures in the main text, we use \prettyref{eq:LIF_Hertz_rescaled}.

From the hazard function, we obtain the firing rate
\begin{equation}
\nu^{-1}=\int_{0}^{\infty}e^{-\int_{0}^{T}h(t)dt}dT\label{eq:LIF_hazard_to_rate}
\end{equation}
as well as the inter--spike--interval distribution \citep{Gerstner14}
\begin{align}
p(T) & =h(T)e^{-\int_{0}^{T}h(t)dt}.\label{eq:LIF_hazard_to_isi_dist}
\end{align}
From the Fourier transform of the inter--spike--interval distribution
$\tilde{p}(f)=\int_{0}^{\infty}e^{2\pi ifT}p(T)dT$, we obtain the
spike-train power spectrum using \citep{Stratonovich67} 
\begin{align}
S_{x}(f) & =\nu\frac{1-|\tilde{p}(f)|^{2}}{|1-\tilde{p}(f)|^{2}}.\label{eq:LIF_isi_dist_to_spec}
\end{align}
Thus, we are left with the task of calculating $n_{1}(t_{1})$ and
$Q(t_{1},t_{2})$.

\subsubsection{Free upcrossing probabilities}

The free voltage dynamics are governed by \prettyref{eq:LIF_voltage}
\begin{align*}
\tau_{\mathrm{{m}}}\dot{U}(t) & =-U(t)+I(t),
\end{align*}
where $I$ is a Gaussian process determined by \prettyref{eq:LIF_current_mean}
and \prettyref{eq:LIF_current_corr}, and the initial condition is
$U(0)=V_{\mathrm{{r}}}$. $U$ is a nonstationary Gaussian process
due to the initial condition. For a sufficiently smooth Gaussian process,
the upcrossing probability is given by the Kac--Rice formulae \citep{Rice45,Stratonovich67,Azais09}
\begin{align*}
n_{1}(t) & =\int_{0}^{\infty}\dot{U}_{1}p(\theta,\dot{U}_{1}\,|\,V_{\mathrm{{r}}},\dot{U}_{0})d\dot{U}_{1},\\
n_{2}(t_{1},t_{2}) & =\int_{0}^{\infty}\int_{0}^{\infty}\dot{U}_{2}\dot{U}_{1}p(\theta,\dot{U}_{2};\theta,\dot{U}_{1}\,|\,V_{\mathrm{{r}}},\dot{U}_{0})d\dot{U}_{1}d\dot{U}_{2},
\end{align*}
where $p(\theta,\dot{U}_{1}\,|\,V_{\mathrm{{r}}},\dot{U}_{0})$ denotes
the probability that the process is at the threshold after time $t$
and has velocity $\dot{U}_{1}$ given that it started at the reset
at $t=0$ with velocity $\dot{U}_{0}$. Similarly, $p(\theta,\dot{U}_{2};\theta,\dot{U}_{1}\,|\,V_{\mathrm{{r}}},\dot{U}_{0})$
denotes the joint probability to be at the threshold at $t_{1}$ and
$t_{2}$ with velocities $\dot{U}_{1}$ and $\dot{U}_{2}$. All integrals
are over positive velocities only, because we consider upcrossings.

In both equations, we need to specify the distribution of the initial
velocity $\dot{U}_{0}$. Here, it is important to take into account
the biased sampling of the initial velocity \citep{Lindner04_0229011}:
at $-\tau_{\mathrm{{ref}}}^{\alpha}$, the neuron spiked due to an
increased input current; hence, the initial velocity $\tau_{\mathrm{{m}}}\dot{U}_{0}=-V_{\mathrm{{r}}}+I_{0}$
is likely to be larger than for an $I_{0}$ drawn from the stationary
current distribution. To keep the integral in \prettyref{eq:LIF_n1}
tractable, we assume that $I_{0}$ is Gaussian-distributed. To determine
the mean and variance of this distribution, we use that the velocity
of a stationary process at an upcrossing is Rayleigh-distributed \citep{Stratonovich67}
(details in Appendix C).

For $n_{2}(t_{1},t_{2})$, we consider only the stationary two-point
upcrossing probability, so that it becomes a function of the time
difference $t_{2}-t_{1}$ and loses the dependency on the initial
velocity. After marginalizing the initial velocity in $n_{1}(t)$,
we obtain

\begin{align}
n_{1}(t) & =\int_{0}^{\infty}\dot{U}_{1}p(\theta,\dot{U}_{1}\,|\,V_{\mathrm{{r}}})d\dot{U}_{1},\label{eq:LIF_n1}\\
n_{2}(t_{2}-t_{1}) & =\int_{0}^{\infty}\int_{0}^{\infty}\dot{U}_{2}\dot{U}_{1}p(\theta,\dot{U}_{2};\theta,\dot{U}_{1})d\dot{U}_{1}d\dot{U}_{2},\label{eq:LIF_n2_stationary}
\end{align}
where $n_{2}(\tau)$ leads to a stationary $Q(\tau)=1-\frac{n_{2}(\tau)}{n_{0}^{2}}$.
This makes the integrals in \prettyref{eq:LIF_Stratonovich_approximation}
considerably easier to solve numerically (details in Appendix D).

Since the free dynamics are linear, $p(\theta,\dot{U}_{1}\,|\,V_{\mathrm{{r}}})$
and $p(\theta,\dot{U}_{2};\theta,\dot{U}_{1})$ can be obtained analytically.
Importantly, the integral in \prettyref{eq:LIF_n1} as well as the
double integral in \prettyref{eq:LIF_n2_stationary} are analytically
solvable using the table \citep{Owen80_389} (details in Appendix B.1
and Appendix C). This is a novel result, to the best of our knowledge,
and considerably simplifies the numerical evaluation of \prettyref{eq:LIF_Hertz_rescaled}.

\paragraph{}

\subsubsection{Numerical solution of the self-consistency problem}

Just as for the GLM networks, we solve the colored noise problem using
a fixed-point iteration. To initiate the algorithm, we set the rates
to $\nu^{\alpha}=1/\tau_{\mathrm{{m}}}^{\alpha}$. We use these rates
to calculate the input mean, variance, and spectrum according to \prettyref{eq:DMFT_mean_pop}
and \prettyref{eq:DMFT_correlation_pop}, beginning with the diffusion
approximation $S_{x}^{\alpha}(t)=\nu^{\alpha}$ and $\sigma_{\nu}^{\alpha}=0$
across neurons. Despite assuming initially equal rates across neurons,
it is possible to have static input variability both due to distributed
indegrees (see \prettyref{eq:DMFT_static} and \prettyref{fig:LIF_BRN_forward}\textbf{a})
and due to evolution of the rates during the fixed-point iteration.
To account for the static variability, we consider an ensemble of
inputs $\mu^{\alpha}+\zeta^{\alpha}$ and determine the corresponding
hazard functions $h_{S}^{\alpha}(t\,|\,\mu^{\alpha}+\zeta^{\alpha})$,
\prettyref{eq:LIF_Hertz_rescaled}, output rates $\nu^{\alpha}(\mu^{\alpha}+\zeta^{\alpha})$,
\prettyref{eq:LIF_hazard_to_rate}, ISI distributions $p^{\alpha}(T\,|\,\mu^{\alpha}+\zeta^{\alpha})$,
\prettyref{eq:LIF_hazard_to_isi_dist}, and spectra $S_{x}^{\alpha}(f\,|\,\mu^{\alpha}+\zeta^{\alpha})$,
\prettyref{eq:LIF_isi_dist_to_spec}. From this ensemble, we obtain
the final output statistics from a numerical average over the ensemble:
\begin{align}
\nu^{\alpha} & =\int_{-\infty}^{\infty}\nu^{\alpha}(\mu^{\alpha}+\zeta^{\alpha})\mathcal{N}(\zeta^{\alpha}\,|\,0,\sigma_{\zeta}^{\alpha})\,d\zeta^{\alpha},\\
({\sigma_{\nu}^{\alpha}})^{2} & =\int_{-\infty}^{\infty}[\nu^{\alpha}(\mu^{\alpha}+\zeta^{\alpha})-\nu^{\alpha}]^{2}\mathcal{N}(\zeta^{\alpha}\,|\,0,\sigma_{\zeta}^{\alpha})\,d\zeta^{\alpha},\\
p^{\alpha}(T) & =\int_{-\infty}^{\infty}p^{\alpha}(T\,|\,\mu^{\alpha}+\zeta^{\alpha})\mathcal{N}(\zeta^{\alpha}\,|\,0,\sigma_{\zeta}^{\alpha})\,d\zeta^{\alpha},\\
S_{x}^{\alpha}(f) & =\int_{-\infty}^{\infty}S_{x}^{\alpha}(f\,|\,\mu^{\alpha}+\zeta^{\alpha})\mathcal{N}(\zeta^{\alpha}\,|\,0,\sigma_{\zeta}^{\alpha})\,d\zeta^{\alpha}.
\end{align}
We solve the above Gaussian integrals using Gauss-Hermite quadrature
\citep{Press07}. Gauss-Hermite quadrature of order $k$ solves Gaussian
integrals of polynomials up to power $k$ exactly by construction.
This allows us to keep the ensemble very small; throughout we use
$k=5$. Finally, we update the statistics using incremental steps,
e.g.~$\nu_{n+1}^{\alpha}=\nu_{n}^{\alpha}+\varepsilon(\hat{\nu}_{n+1}^{\alpha}-\nu_{n}^{\alpha})$
for the firing rate, where $\hat{\nu}_{n+1}^{\alpha}$ denotes the
estimated rate based on the input at the previous step. Here, the
small update step $\varepsilon<1$ is crucial because otherwise the
algorithm is numerically unstable. Now we iterate and generate new
input statistics. Repeated application of this scheme suggests that
the self-consistent problem for the type of networks under consideration
possesses only a single fixed point to which the algorithm always
converges.

\subsection{Balanced random network}

First, we consider the same balanced random network as we did for
the GLM neurons (\prettyref{fig:BRN-GLM-EXP}\textbf{a}). In particular,
we place the network in the inhibition-dominated regime, drive the
network with a constant external input, and use identical single-neuron
parameters for excitatory and inhibitory neurons. In order to obtain
a biologically plausible activity below $10\:\mathrm{spks/s}$, we
keep the external input weak to place the network deep in the fluctuation-driven
regime. In this regime, the mean input to a neuron is far below threshold
and only occasional large fluctuations in the input drive it above
the spike threshold (\prettyref{fig:LIF_BRN_forward}\textbf{a},\textbf{b}).
If the mean inter-spike interval exceeds the correlation time of the
input, the renewal approximation is admissible. Indeed, since the
firing rates are low by construction, even moderate input correlation
times are smaller than the inverse firing rate.

\subsubsection{Colored noise problem}

\begin{figure}
\includegraphics{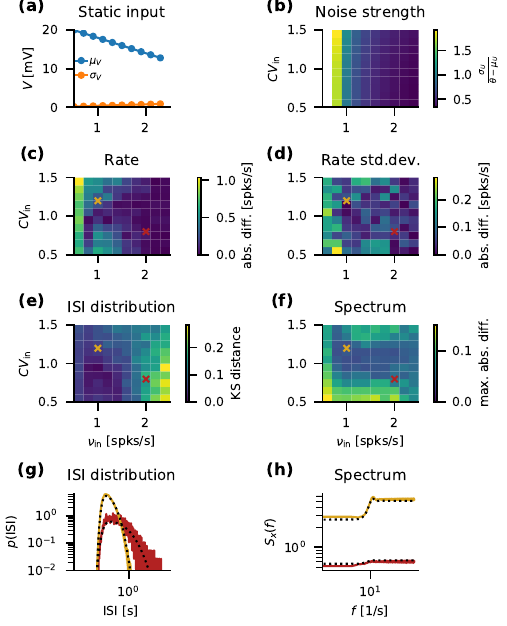}

\caption{Colored noise problem in a balanced random network of LIF neurons.
Comparison between theory, \prettyref{eq:LIF_Hertz_rescaled}, and
LIF neurons driven by Gaussian processes (GPs). \textbf{(a)} Mean
(orange) and standard deviation across neurons (blue) of the membrane
potential due to the static contribution $\tau_{\mathrm{{m}}}\zeta$.
\textbf{(b)} Noise strength of the effective input measured by the
standard deviation of the membrane potential fluctuations relative
to the distance to threshold $\sigma_{U}/(\theta-\mu_{U})$. \textbf{(c)}
Absolute difference between rate from theory and GP-driven LIF neurons.
\textbf{(d)} Same as (c) but for the standard deviation of the rate
across neurons. \textbf{(e)} Kolmogorov-Smirnov distance using $2.5\,\mathrm{ms}$
bins between ISI distribution from theory and GP-driven LIF neurons.
\textbf{(f)} Maximal absolute distance between power spectra from
theory and GP-driven LIF neurons. \textbf{(g},\textbf{h)} Example
ISI distributions and power spectra from theory (black) and GP-driven
LIF neurons (colored) for the parameter values indicated by crosses
in (c--f). Parameters: $N_{E}=40,\!000$, $N_{I}=10,\!000$, $J_{E}=0.1\,\mathrm{mV}$,
$|J_{I}/J_{E}|=6.0$, $p=0.1$, $\tau_{\mathrm{m}}=20\,\mathrm{ms}$,
$\tau_{\mathrm{s}}=5\,\mathrm{ms}$, $\tau_{\mathrm{ref}}=2\,\mathrm{ms}$,
$d=1.5\,\mathrm{ms}$, $\theta=20\,\mathrm{mV}$, $V_{\mathrm{r}}=0\,\mathrm{mV}$,
$\mu_{\mathrm{ext}}=22\,\mathrm{mV}$. \label{fig:LIF_BRN_forward}}
\end{figure}

First, we isolate the colored noise problem to gauge the above approximations.
To this end, we examine a single step in the fixed-point iteration.
For this single step, we compare the theory with a population of unconnected
LIF neurons driven by independent Gaussian processes (GPs). A self-consistent
solution using populations of LIF neurons driven by GPs, where the
colored noise problem is thus solved numerically, accurately captures
the network-averaged power spectrum of the single-unit activity of
the connected network \citep{dummer14}. In particular, this implies
that cross-correlations as well as the non-Gaussianity of the input
statistics can be neglected, in agreement with the DMFT prediction.
Hence, if our theory yields results comparable to a population of
GP-driven LIF neurons for a single step of the fixed-point iteration,
the self-consistent fixed point will capture the simulation well.
We compare the self-consistent theory with simulations in the following
sections.

We do not determine the effective input statistics using simulation
results here, because this would preclude a systematic scan over the
parameters of the input, which consists of both external and recurrent
network contributions. Instead, we fix the effective external input
and determine the statistics of the effective recurrent input in terms
of the input spiking statistics $\nu_{\mathrm{in}}$, $\sigma_{\nu}^{\mathrm{in}}=0$
across neurons, and
\begin{equation}
S_{x}^{\mathrm{in}}(f)=\nu_{\mathrm{in}}\frac{1-|(1-2\pi i\mathrm{CV}_{\mathrm{in}}^{2}f/\nu_{\mathrm{in}})^{-1/\mathrm{CV}_{\mathrm{in}}^{2}}|^{2}}{|1-(1-2\pi i\mathrm{CV}_{\mathrm{in}}^{2}f/\nu_{\mathrm{in}})^{-1/\mathrm{CV}_{\mathrm{in}}^{2}}|^{2}},\label{eq:spectrum_gamma}
\end{equation}
corresponding to a gamma process with rate $\nu_{\mathrm{in}}$ and
CV of the ISI distribution $\mathrm{CV}_{\mathrm{in}}$, cf.~\prettyref{eq:LIF_isi_dist_to_spec}.
This leaves a two-dimensional parameter space spanned by $\nu_{\mathrm{in}}$
and $\mathrm{CV}_{\mathrm{in}}$. From the spiking statistics, we
obtain the statistics of the effective input using \prettyref{eq:DMFT_mean_pop}
and \prettyref{eq:DMFT_correlation_pop}. Note that although $\sigma_{\nu}^{\mathrm{in}}=0$,
the static variability of the effective input is nonzero, $\sigma_{\zeta}>0$,
due to the distributed indegree, see \prettyref{eq:DMFT_static} and
\prettyref{fig:LIF_BRN_forward}\textbf{a}. Hence, we can compare
both the averaged output statistics and the rate variability in the
population. For the comparison, we simulate $250$ GP-driven LIF neurons
for $50\,\mathrm{s}$ with a time step of $0.05\,\mathrm{ms}$; we
use the same interval and time step for the theory.

Guided by the regime attained in full simulations, we choose $\nu_{\mathrm{in}}\in[0.5,2.5]\,\mathrm{spks/s}$
and $\mathrm{CV}_{\mathrm{in}}\in[0.5,1.5]$ (\prettyref{fig:LIF_BRN_forward}).
The network is in the inhibition-dominated regime; thus the mean input
decreases with $\nu_{\mathrm{in}}$ starting from the value that brings
the membrane potential on average to threshold (\prettyref{fig:LIF_BRN_forward}\textbf{a}).
In contrast, the static variability increases monotonically with $\nu_{\mathrm{in}}$
(\prettyref{fig:LIF_BRN_forward}\textbf{a}). To measure the strength
of the dynamic variability, we divide the resulting standard deviation
of the free membrane voltage by the distance of the mean free membrane
voltage to the threshold, $\sigma_{U}/(\theta-\mu_{U})$. Since the
numerator grows with $\sqrt{\nu_{\mathrm{in}}}$ while the denominator
grows linearly with $\nu_{\mathrm{in}}$ in inhibition-dominated networks,
the standard deviation relative to the distance to threshold decreases
with increasing $\nu_{\mathrm{in}}$; in contrast, it slightly increases
with increasing $\mathrm{CV}_{\mathrm{in}}$ (\prettyref{fig:LIF_BRN_forward}\textbf{b}).
For the entire parameter regime, the absolute difference in the firing
rate is smaller than $1\,\mathrm{spks/s}$ and it is maximal at the
brink of the fluctuation-driven regime (\prettyref{fig:LIF_BRN_forward}\textbf{c}).
For the static rate variability, we also consider the absolute difference,
which is below $0.3\,\mathrm{spks/s}$ throughout the parameter space
(\prettyref{fig:LIF_BRN_forward}\textbf{d}). Next, we compare the
ISI distributions using their Kolmogorov-Smirnov distance, i.e.~the
maximal absolute difference between the cumulative distributions.
The Kolmogorov-Smirnov distance is maximal deep in the fluctuation-driven
regime where the firing rate is well below $1\,\mathrm{spks/s}$ and
the estimate of the ISI distribution is noisy (\prettyref{fig:LIF_BRN_forward}\textbf{e}).
Finally, we compare the output spectra using the maximum absolute
distance between the scaled spectra $S_{x}(f)/\nu$. Here, the deviation
is below $0.1$ in most parts of the parameter space except for low
$\mathrm{CV_{\mathrm{in}}}\le0.6$, high $\mathrm{CV}_{\mathrm{in}}\ge1.3$,
and at the brink of the fluctuation-driven regime (\prettyref{fig:LIF_BRN_forward}\textbf{f}).
To give meaning to the quantitative results, we plot two example ISI
distributions (\prettyref{fig:LIF_BRN_forward}\textbf{g}) and spectra
(\prettyref{fig:LIF_BRN_forward}\textbf{h}). For the ISI distribution,
we see the noisy estimate at low rates. For the spectra, we note that
the main difference is a constant offset caused by a small error in
the rate, see \prettyref{eq:LIF_isi_dist_to_spec}, while the shape
is well matched.

To conclude, the above approximations work well in the fluctuation-driven
regime for moderate values $\mbox{\ensuremath{0.6<\mathrm{CV}_{\mathrm{in}}<1.3}}$.
Within this regime, the firing rate and its variability across neurons,
the ISI distribution, and the power spectra are well predicted. Most
importantly for the prediction of the intrinsic timescale, the theory
closely predicts the scaled spectrum $S_{x}(f)/\nu$.

\subsubsection{Timescales in balanced random networks of LIF neurons}

\begin{figure}
\centering{}\includegraphics{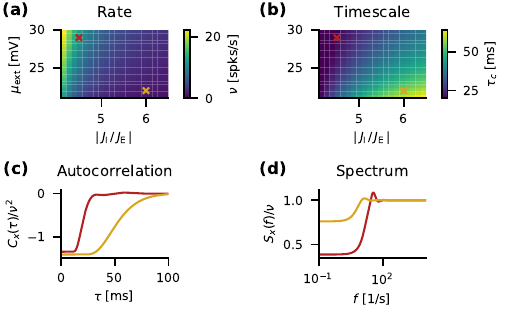}\caption{Parameter scan for a balanced random network of LIF neurons using
\prettyref{eq:LIF_Hertz_rescaled}. \textbf{(a},\textbf{b)} Firing
rate and intrinsic timescale for varying external input $\mu_{\mathrm{ext}}$
and relative inhibitory strength $|J_{I}/J_{E}|$. \textbf{(c},\textbf{d)}
Scaled autocorrelation $C_{x}(\tau)/\nu^{2}$ and power spectrum $S_{x}(f)/\nu$
for the parameter values indicated by symbols in (a,b). Further parameters
as in \prettyref{fig:LIF_BRN_forward}. \label{fig:LIF_BRN_scan}}
\end{figure}

Having established the validity of the theory, we employ it to investigate
the intrinsic timescale. It is well known that increasing the overall
synaptic strength leads to a network state with long temporal correlations
\citep{Ostojic14,Wieland2015_040901}. However, this state comes along
with giant fluctuations of the membrane potential \citep{Kriener2014_136}
which are well beyond the physiological regime and which our theory
can capture only to a limited extent. Hence, we focus on the influence
of the external input $\mu_{\mathrm{ext}}$ and the inhibition dominance
$|J_{I}/J_{E}|$, in line with our above investigations for GLM neurons.
We solve the theory on a $\Delta t=0.05\,\mathrm{ms}$ grid to a maximum
of $T=10\,\mathrm{s}$, use an ensemble size of $k=5$ for the Gauss-Hermite
quadrature, and choose an update step $\varepsilon=0.2$.

We investigate the regime $|J_{I}/J_{E}|\in[4.1,6]$ and $\mbox{\ensuremath{\mu_{\mathrm{ext}}\in[21,30]\,\mathrm{mV}}}$.
Within this regime, the rate is below approximately $20\,\mathrm{spks/s}$,
increases with $\mu_{\mathrel{\mathrm{ext}}}$, and decreases with
$|J_{I}/J_{E}|$ (\prettyref{fig:LIF_BRN_scan}\textbf{a}). In contrast,
the intrinsic timescale decreases with $\mu_{\mathrel{\mathrm{ext}}}$,
increases with $|J_{I}/J_{E}|$, and reaches a maximum of approximately
$60\,\mathrm{ms=3\tau_{\mathrm{m}}}$ (\prettyref{fig:LIF_BRN_scan}\textbf{b}).
The autocorrelation function reveals that the nature of these longer
intrinsic timescales in LIF networks is fundamentally different to
the GLM networks above (\prettyref{fig:LIF_BRN_scan}\textbf{c}):
in the GLM networks, the autocorrelation function is positive, which
corresponds to an increased probability to spike in succession; in
the LIF networks it is negative, which corresponds to a prolonged
effective refractory period caused by the fire-and-reset mechanism
in combination with the input statistics. Indeed, in the corresponding
power spectra and their zero-frequency limit, we see that the $\mathrm{CV}$
is well below $1$ (\prettyref{fig:LIF_BRN_scan}\textbf{d}). Hence,
the process is more regular than a Poisson process, as opposed to
the high irregularity $\mathrm{CV}>1$ that would go along with bursty
spiking.

\subsubsection{Simulation of balanced random network of LIF neurons}

\begin{figure}
\includegraphics{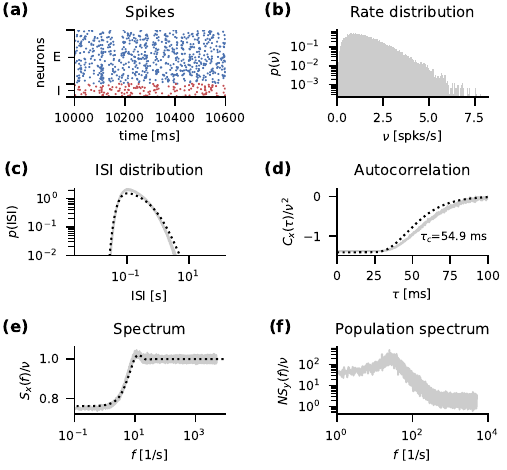}

\caption{Balanced random network of LIF neurons. \textbf{(a)} Raster plot of
2\% of the excitatory (blue) and inhibitory (red) neurons. \textbf{(b)}
Firing rate distribution across all neurons. \textbf{(c},\textbf{d},\textbf{e)}
Population-averaged ISI distribution, population-averaged autocorrelation
function, and population-averaged power spectrum from simulation (gray)
and theory (black). \textbf{(f)} Power spectrum of the population
activity. Parameters as in \prettyref{fig:LIF_BRN_forward}.\label{fig:LIF_BRN_network}}
\end{figure}

We validate the theoretical predictions for the balanced random network
of LIF neurons by comparing with a network simulation. To acquire
sufficient statistics, we simulate the network for $T=2.5\,\mathrm{min}$
with time step $\Delta t=0.1\,\mathrm{ms}$ and discard the first
$1\,\mathrm{s}$ as an initial transient. After this transient, the
network is in an asynchronous irregular state (\prettyref{fig:LIF_BRN_network}\textbf{a}).
The rates of individual neurons are mostly below $5\,\mathrm{spks/s}$
with a peak at around $1\,\mathrm{spks/s}$ (\prettyref{fig:LIF_BRN_network}\textbf{b}).
The theory closely predicts the ISI distribution apart from a slight
overestimation of the tail (\prettyref{fig:LIF_BRN_network}\textbf{c}).
Thus, the resulting autocorrelation function is also well matched
and the predicted intrinsic timescale of approximately $55\,\mathrm{ms}$
is confirmed (\prettyref{fig:LIF_BRN_network}\textbf{d}). Also the
scaled spectrum is closely reproduced and reveals a $\mathrm{CV}^{2}\approx0.75$
(\prettyref{fig:LIF_BRN_network}\textbf{e}).

To illustrate the difference between the single-unit and the population
statistics, we furthermore plot the power spectrum of the population
activity $y(t)=\frac{1}{N}\sum_{i=1}^{N}x_{i}(t)$ (\prettyref{fig:LIF_BRN_network}\textbf{f}).
For vanishing cross-correlations, these two spectra would be proportional
to each other. Already weak cross-correlations can shape the population
spectrum since their contribution is of $O(N^{2})$ compared to $O(N)$
contributions from the autocorrelations, leading to the clear differences
we see between the single-unit and the population spectrum. A notable
difference between the two spectra is the peak around $30\:\mathrm{Hz}$
in the population spectrum, contrasting with the roughly $10\:\mathrm{Hz}$
peak in the single-unit spectrum. Furthermore, the population spectrum
displays increased power at low frequencies compared to high frequencies,
while the reverse is true for the single-unit spectrum.

\subsection{Biologically Constrained Network Model}

\begin{figure}
\centering{}\includegraphics[width=1\columnwidth]{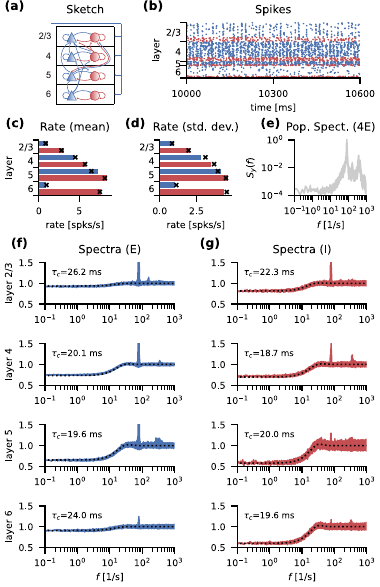}\caption{Multi-population network of LIF neurons. \textbf{(a) }Sketch of the
model, figure adapted from \citep{Potjans14_785}. \textbf{(b)} Raster
plot of 2\% of the neurons of each population. \textbf{(c},\textbf{d)}
Neuron-averaged firing rates and their standard deviation from simulations
(bars) and theoretical predictions (black crosses). \textbf{(e)} Population
spectrum of the layer 4 excitatory population. \textbf{(f},\textbf{g)}
Spike-train power spectra obtained from simulations (colored) and
theory (black) and the corresponding intrinsic timescale. Parameters
as specified in \citep{Potjans14_785}. \label{fig:LIF_microcircuit}}
\end{figure}

Thus far, we only considered balanced random networks with identical
excitatory and inhibitory neurons that reduce to a single effective
population. Despite this simplification, these balanced random networks
already span a large parameter space. Here, we apply our theory to
a multi-population network model constrained by biological data \citep{Potjans14_785}.
Beyond the aspect of multiple populations, this network model allows
us to highlight two additional features of our theory that we left
out thus far: the possibility to include external Poisson input and
distributed synaptic weights. We solve the theory on a $\Delta t=0.05\,\mathrm{ms}$
grid to a maximum of $T=10\,\mathrm{s}$, use an ensemble size of
$k=5$ for the Gauss-Hermite quadrature, choose an update step $\varepsilon=0.1$,
and initialize all populations with a rate of $10\,\mathrm{spks/s}$.

The model represents the neurons under $1\,\mathrm{mm}^{2}$ of surface
of generic early sensory cortex. It comprises eight populations: layers
2/3, 4, 5, and 6 with a population of excitatory cells and inhibitory
interneurons for each layer (\prettyref{fig:LIF_microcircuit}\textbf{a}).
In total, this leads to $77,\!169$ neurons connected via approximately
$3\times10^{8}$ synapses, with population-specific connection probabilities
$p^{\alpha\beta}$ based on an extensive survey of the anatomical
and physiological literature. In contrast to the original model, we
directly use the connection probabilities to create the connectivity
such that the total number of synapses can vary across instantiations
of the model, and we draw source and target neurons without replacement,
so that multapses are not allowed. Transmission delays follow truncated
normal distributions with mean \textpm{} standard deviation of $1.5\pm0.75\:\mathrm{ms}$
for excitatory source neurons and $0.75\pm0.375\:\mathrm{ms}$ for
inhibitory source neurons, both with a cutoff at $0.1\:\mathrm{ms}$.
The synaptic strengths $\synmatrix_{ij}^{\alpha\beta}$ are normally
distributed with $\mu_{\synmatrix}^{\alpha I}=-351.2\:\mathrm{pA}$
for inhibitory source neurons and $\mu_{\synmatrix}^{\alpha E}=87.8\:\mathrm{pA}$
for excitatory source neurons except for connections from layer 4
excitatory to layer 2/3 excitatory neurons, which have a mean strength
of $175.6\:\mathrm{pA}$. For all synaptic strengths, the standard
deviation is fixed to $10\%$ of the mean. The network is driven by
external Poisson input with layer-specific rates (for further details
see \citep{Potjans14_785}).

The intrinsic parameters of the neurons do not vary across populations.
Shaped by the connectivity, a layer-specific activity arises (\prettyref{fig:LIF_microcircuit}\textbf{b})
with mean firing rates between $1$ and $10\:\mathrm{spks/s}$ (\prettyref{fig:LIF_microcircuit}\textbf{c})
and a standard deviation across neurons between $1$ and $5\:\mathrm{spks/s}$
(\prettyref{fig:LIF_microcircuit}\textbf{d}). While the quantitative
agreement is not perfect, our theory captures the specificity of both
mean firing rate and its variability across neurons well.

A prominent feature of the model are oscillations on the population
level \citep{Bos16_1} which are already visible in the raster plot
of only 2\% of the population (\prettyref{fig:LIF_microcircuit}\textbf{b}).
These oscillations lead to a clear peak at about $80\:\mathrm{Hz}$
in the power spectrum of the population activity in all layers \citep{Bos16_1}.
Here, we only show a representative population spectrum (\prettyref{fig:LIF_microcircuit}\textbf{e}).
These population-level oscillations clearly violate the independence
assumption of the effective inputs. Thus, they could potentially explain
the deviations of the predicted firing rate from the simulation.

For most populations, the peak in the population-level oscillations
also manifests itself in the population-averaged single-unit spectra
(\prettyref{fig:LIF_microcircuit}\textbf{f},\textbf{g}). Apart from
this peak, our theory closely captures the shape of all spectra (\prettyref{fig:LIF_BRN_forward}\textbf{f},\textbf{g}).
Note that, despite the large heterogeneity of mean rates, the intrinsic
timescale is similar across populations. As in the balanced random
network, the intrinsic timescales are on the order of magnitude of
the membrane time constant (here $10\:\mathrm{ms}$); concretely,
the intrinsic timescale is approximately twice as large.

\section{Discussion}

We developed a self-consistent theory for the second-order statistics,
in particular the intrinsic timescales as defined by autocorrelation
decay times, in block-structured random networks of spiking neurons.
Orthogonal to approaches based on the mean activity of a population
of neurons, we consider population-averaged single-neuron statistics.
To this end, we built on the model-independent dynamic mean-field
theory (DMFT) developed in \citep{Keup21_021064} and applied it to
networks of spiking neurons. We sketched the derivation starting from
the characteristic functional of the recurrent input, \prettyref{eq:charfct_input},
to expose the inherent assumptions of the DMFT as well as its main
result. In particular, we showed that the mean-field equations, \prettyref{eq:DMFT_mean_pop}
and \prettyref{eq:DMFT_correlation_pop}, where the connectivity matrix
enters only through its first two cumulants, account for both (static)
inter-neuron variability and (dynamic) temporal fluctuations. In order
to close the self-consistency problem, we derived a novel analytical
solution for the output statistics of a generalized linear model (GLM)
neuron with error-function nonlinearity driven by a Gaussian process
(GP), \prettyref{eq:GLM_erf_corr}, and an analytical approximation
for the output statistics of a GP-driven leaky integrate-and-fire
(LIF) neuron in the fluctuation-driven regime, \prettyref{eq:LIF_Hertz_rescaled}.
These theoretical results yield firing rate distributions, spike-train
power spectra, and inter-spike interval distributions that are close
to those obtained from numerical simulations (\prettyref{fig:BRN-GLM-EXP},
\prettyref{fig:BRN-GLM-ERF}, \prettyref{fig:LIF_BRN_network}) even
for a complex, biologically constrained network model (\prettyref{fig:LIF_microcircuit}).

The excellent agreement between theory and simulations demonstrates
the validity of the DMFT approximation, i.e.~the approximation of
the recurrent inputs as independent Gaussian processes. The validity
of the DMFT approximation is most clearly demonstrated by the networks
of GLM neurons, since in that case the DMFT assumption constitutes
the only approximation, while the remainder of the solution is exact;
while for the LIF networks, additional approximations are made, so
that the effects of the DMFT assumption can be less well isolated.

Focusing on balanced random networks, we leveraged our theory to investigate
the influence of network parameters on the intrinsic timescale for
both GLM (\prettyref{fig:GLM-EXP-SCAN}, \prettyref{fig:GLM-ERF-SCAN})
and LIF (\prettyref{fig:LIF_BRN_scan}) neurons. For the former neuron
model with error function nonlinearity, our theory unveils that a
product of two factors determines the intrinsic timescale (\prettyref{eq:GLM_tau_inf},
\prettyref{fig:GLM_ERF_mechanism}): the gain of the rate autocorrelation
function with respect to changes in the membrane voltage autocorrelation
function for $\tau\to\infty$, \prettyref{eq:GLM_erf_gain}, and the
variance of the connectivity, \prettyref{eqs:g_g_bar}. Furthermore,
providing a temporally correlated external drive causes the intrinsic
timescale to monotonically approach the extrinsic timescale as the
input strength is increased (\prettyref{fig:GLM_ERF_tauext}).

For both GLM neurons with error function nonlinearity and LIF neurons,
we find parameter regimes where the intrinsic timescale $\tau_{c}$
is longer than the largest time constant of the single-neuron dynamics,
the membrane time constant $\tau_{\mathrm{m}}$ (\prettyref{fig:BRN-GLM-ERF},
\prettyref{fig:LIF_BRN_scan}). This demonstrates that the recurrent
dynamics shape the intrinsic timescale. Note that we consider a regime
where the inverse firing rate $\nu^{-1}$ is large compared to $\tau_{\mathrm{m}}$.
In contrast, \citep{Kadmon15_041030,Harish15_e1004266} consider the
opposite regime where slow neuronal timescales lead to effective rate
dynamics, and the spiking noise is either left out or treated perturbatively.
Our results show that it is possible to obtain longer intrinsic timescales
even in a regime where the white component of the spiking noise contributes
non-negligibly to the membrane voltage fluctuations. However, the
temporal structure that causes the prolonged intrinsic timescale is
very different for the two models that we consider: For GLM neurons,
the autocorrelation is positive for a period on the order of $\tau_{c}$,
corresponding to an increased spiking probability. For LIF neurons,
the autocorrelation function is negative, corresponding to a prolonged
effective refractory period.

Furthermore, LIF networks exhibit a minimum in the intrinsic timescale
\citep{Wieland2015_040901} while the corresponding GLM networks exhibit
a maximum (\prettyref{fig:GLM_ERF_mechanism}). We hypothesize that
this difference is due to the difference in the temporal structure:
The minimum in the timescale for LIF networks is caused by a switch
from an increased effective refractory period (a negative autocorrelation
function for $\tau\to0$) to an increased probability for another
spike (a positive autocorrelation function for $\tau\to0$). For GLM
networks, this switch and hence the minimum is absent. Instead, the
more subtle interplay between the gain and the variance of the connectivity
leads to the maximum. The presence of a maximum rather than a minimum
in the intrinsic timescales renders the GLM networks more similar
to networks of rate units \citep{Sompolinsky88_259}. If similar mechanisms
are at play as in rate networks, the white spiking noise of the input
to the GLM neurons may temper the size of the largest possible timescale
\citep{Schuecker18_041029}. However, due to the inherent stochasticity
of GLM neurons, it is unclear whether the maximum occurs at a transition
to chaos as it does in rate networks \citep{Sompolinsky88_259}. 

Considering a more complex block-structured network model that is
constrained by biological data \citep{Potjans14_785} exposes limits
of our theory: while the theory accurately captures the non-oscillatory
components of the power spectra, it misses a high-frequency oscillation
(\prettyref{fig:LIF_microcircuit}). These high-frequency oscillations
are caused by correlated activity on the population level \citep{Bos16_1};
hence, the peak in the population-averaged single-neuron spectra demonstrates
an interplay between single-unit and population-level statistics that
was absent in the simpler balanced random network models. By construction,
our theory only accounts for population-averaged single-neuron statistics
and thus misses the high-frequency peak. It is an interesting challenge
to derive a self-consistent theory on both scales simultaneously.

In general, the limits of DMFT when applied to spiking networks merit
further investigation. For example, assuming that the network is sparse,
$K\ll N$ or $p\ll1$, is not a necessary condition for a DMFT to
apply \citep{Kadmon15_041030}. Nonetheless, increasing sparsity reduces
the pairwise correlations between the neurons \citep{Vreeswijk98_1321,Brunel00_183,Ostojic14}
such that DMFT is expected to yield better results. Another important
aspect is that for the synaptic weights scaling as $J_{ij}=O(1/\sqrt{K})$,
the fluctuations of the mean input $\mu_{\eta}(t)$ can be $O(1)$,
i.e., not scale with $K^{-\alpha}$, $\alpha>0$, as the network size
increases and $p$ is kept constant. In \prettyref{eqs:DMFT_1pop},
$\mu_{\eta}(t)$ and $C_{\eta}(t,t^{\prime})$ are replaced by their
average, neglecting fluctuations; including the fluctuations of the
mean input would lead to an additional term in $C_{\eta}(t,t^{\prime})$
\citep{Helias14}. Since these fluctuations of the mean input reflect
pairwise correlations, the latter need to be small for the theory
to be accurate. The above scaling argument shows that it is nontrivial
that the pairwise correlations vanish, even in the large network limit.
They only do so for an asynchronous state in which the pairwise correlations
are small already for finite networks, e.g., due to a sparse network
or due to inhibitory feedback \citep{Tetzlaff12_e1002596}. Conversely,
if a network is in an asynchronous irregular state, which has low
pairwise correlations by definition, DMFT is expected to yield reliable
results.

The heterogeneity of timescales even within a cortical area \citep{Cavanagh20_81}
suggests another interesting extension, namely to calculate the variability
of the timescale within a population. This requires calculating the
variability of the second-order statistics, which has recently been
achieved for linear rate networks \citep{Dahmen19_13051} but to the
best of our knowledge is an open challenge even for simple nonlinear
rate networks, let alone for spiking networks.

The microscopic theory presented here enables direct comparisons with
experimental measurements of neuron-level intrinsic timescales \citep{Murray14},
in contrast to previous works which have considered population rate
models \citep{Chaudhuri14_e01239,Chaudhuri2015_419}. It is important
to distinguish between neuron-level and population-level autocorrelations,
since the latter are shaped by $O(N^{2})$ cross-correlations and
can therefore differ substantially from neuron-level autocorrelations,
as we have illustrated for the balanced random network model (\prettyref{fig:LIF_BRN_network}\textbf{e},\textbf{f})
and the biologically constrained network model (\prettyref{fig:LIF_microcircuit}\textbf{e}-\textbf{g}).

Establishing a direct link between the connectivity and the emergent
intrinsic timescales opens up the possibility of a thorough investigation
of the effect of network architecture. Moreover, within our theory,
it is possible to account for population-specific intrinsic neuron
parameters. Thus, the theory also provides an avenue for investigations
of the complex interplay between intrinsic parameters \citep{Duarte17_156,Wang20_169}
and the network structure \citep{Chaudhuri2015_419}. In this context,
an interesting application is clustered networks which feature slow
switching between transiently active clusters \citep{LitwinKumar12_1498}.
In particular, clustered networks with both excitatory and inhibitory
clusters \citep{Rost18_81,Rostami20_bioRxiv,Kim21_129} could be of
interest because they robustly give rise to winnerless competition.
From a modeler's point of view, uncovering mechanisms shaping intrinsic
timescales could be used to fine-tune network models \citep{Deco12_3366,Garagnani17_145,Tomasello18_88,Schmidt18_1409}
to match the experimentally observed hierarchy of timescales \citep{Murray14}.
Focusing on computational aspects, diverse timescales strongly enhance
the computational capacity of a recurrent network \citep{Sussillo14_156,Barak17_1,Bellec18_795},
and neurons with long intrinsic timescales carry more information
in a working memory task \citep{Wasmuht18_1} (but see \citep{Spitmaan20_20}).
In this light, the results presented here may also contribute to improved
understanding of aspects of information processing in the brain.

\paragraph*{This work was supported by the European Union's Horizon 2020 Framework
Programme for Research and Innovation under Specific Grant Agreements
No. 785907 and 945539 (Human Brain Project SGA2, SGA3), the Jülich-Aachen
Research Alliance (JARA), and DFG Priority Program \textquotedblleft Computational
Connectomics\textquotedblright{} (SPP 2041). AvM would like to thank
Moritz Helias for many helpful and inspiring discussions about dynamical
mean-field theory and spiking neurons, Tilo Schwalger for his insights
about the Stratonovich approximation and for sharing his unpublished
manuscript, and Jasper Albers, Anno Kurth, Alessandra Stella, Christian
Keup, and David Dahmen for valuable comments on an early version of
the manuscript.}

\section*{Appendix}

\subsection{Characteristic Functionals\label{Appendix A}}

Here, we briefly introduce the characteristic functionals for both
types of stochastic processes we consider: Gaussian processes and
point processes. We closely follow Stratonovich's book \citep{Stratonovich67},
in particular chapters I.1.\ and I.6.

\subsubsection{Stochastic Processes}

The characteristic functional of a stochastic process $\xi(t)$ is
defined as
\begin{align*}
\Phi_{\xi}[u(t)] & =\left\langle e^{i\int_{0}^{T}u(t)\xi(t)dt}\right\rangle _{\xi}.
\end{align*}
In terms of the cumulants $k_{r}(t_{1},\dots,t_{r})$ the characteristic
functional can be written as 
\begin{align}
\Phi_{\xi}[u(t)] & =e^{\sum_{s=1}^{\infty}\frac{i^{s}}{s!}\int_{0}^{T}\dots\int_{0}^{T}k_{s}(t_{1},\dots,t_{s})u(t_{1})\dots u(t_{s})dt_{1}\dots dt_{s}}.
\end{align}
All properties of a stochastic process are determined by its characteristic
functional.

If all cumulants except for the first vanish, the process is deterministic
and has the characteristic functional
\begin{align}
\Phi_{\xi}[u(t)] & =\left\langle e^{i\int_{0}^{T}u(t)\xi(t)dt}\right\rangle _{\xi}=e^{i\int_{0}^{T}u(t)\xi(t)dt}.\label{eq:appendix_charfctl_deterministic}
\end{align}
In this case, the first cumulant coincides with the process itself,
$k_{1}(t)=\xi(t)$. If only the first and the second cumulants are
non-vanishing, the process is a Gaussian process. The corresponding
characteristic functional reads 
\begin{align}
\Phi_{\xi}[u(t)] & =e^{i\int k_{1}(t_{1})u(t_{1})dt_{1}-\tfrac{1}{2}\int\int u(t_{1})k_{2}(t_{1},t_{2})u(t_{2})dt_{1}dt_{2}}.\label{eq:appendix_charfctl_GP}
\end{align}
If the Gaussian process is stationary, $k_{1}(t_{1})=k_{1}$ and $k_{2}(t_{1},t_{2})=k_{2}(t_{2}-t_{1})$,
the characteristic functional simplifies further to $\Phi_{\xi}[u(t)]=e^{ik_{1}\int u(t_{1})dt_{1}-\tfrac{1}{2}\int\int u(t_{1})k_{2}(t_{2}-t_{1})u(t_{2})dt_{1}dt_{2}}$.

The characteristic functional describes the statistics at all points
in time.  It is often useful to relate the characteristic functional
to the distribution of the values of $\xi(t)$ at fixed points in
time, for instance to compute the statistics of the current at upcrossings
and after the refractory period, or to obtain marginal activity statistics
which, given stationarity, reflect time-averaged activity. To this
end, we can use the test functions $u(t)=u_{1}\delta(t-t_{1})$ and
$u(t)=u_{1}\delta(t-t_{1})+u_{2}\delta(t-t_{2})$ to obtain 
\begin{align}
\Phi_{\xi}(u_{1}) & =e^{ik_{1}(t_{1})u_{1}-\tfrac{1}{2}k_{2}(t_{1},t_{1})u_{1}^{2}},\label{eq:appendix_charfctl_gauss_1}\\
\Phi_{\xi}(u_{1},u_{2}) & =e^{i(k_{1}(t_{1})u_{1}+k_{1}(t_{2})u_{2})}\nonumber \\
 & \times e^{-\tfrac{1}{2}(k_{2}(t_{1},t_{1})u_{1}^{2}+2k_{2}(t_{1},t_{2})u_{1}u_{2}+k_{2}(t_{2},t_{2})u_{2}^{2})}.\label{eq:appendix_charfctl_gauss_2}
\end{align}
These are the characteristic functions of a Gaussian with cumulants
determined by $k_{1}$ and $k_{2}$. Knowing these characteristic
functions for all times $t_{1}$ and $t_{2}$ provides the full picture;
this is the marginalization property of Gaussian processes \citep{Williams06}.

\subsubsection{Point Processes}

The equivalent to the characteristic functional for a point process
is the generating functional. For a spike train $\{t_{1},\dots,t_{n}\}$
(a ``system of random points'' in Stratonovich's naming) with $t_{i}\in[0,T]$
for all $i$, the generating functional is defined by 
\begin{align*}
L_{T}[v(t)] & =\left\langle \prod_{j=1}^{n}[1+v(t_{j})]\right\rangle .
\end{align*}
Here, the number of spikes $n$ is itself a random variable because
the average is taken with respect to all possible realizations of
the spike train \citep{Kuznetsov65_101}.

For point processes, the role of the moments is taken by the ``distribution
functions'' $n_{r}(t_{1},\dots,t_{r})$ which denote the probability
of having at least one point in each interval $[t_{i},t_{i}+dt]$.
The role of the cumulants is taken by the functions $g_{r}(t_{1},\dots,t_{r})$
which are related to the distribution functions as the cumulants of
a stochastic process are related to its moments. In terms of the $g_{r}(t_{1},\dots,t_{r})$,
the generating functional can be written as \citep{Kuznetsov65_101}
\begin{align}
L_{T}[v(t)] & =e^{\sum_{s=1}^{\infty}\frac{1}{s!}\int_{0}^{T}\dots\int_{0}^{T}g_{s}(t_{1},\dots,t_{s})v(t_{1})\dots v(t_{s})dt_{1}\dots dt_{s}}.
\end{align}
The generating functional is directly related to a few useful quantities:
The characteristic function of the number of spikes $n$ in the interval
$[0,T]$ is given by $\langle e^{inu}\rangle=L_{T}[e^{iu}-1]$; the
probability that no point falls into $[0,T]$, i.e.~the survival
probability, is given by $L_{T}[-1]$. The simplest case of a point
process where only $g_{1}$ is non-vanishing is a Poisson process.
The corresponding generating functional reads 
\begin{align*}
L_{T}[v(t)] & =\exp\left(\int_{0}^{T}g_{1}(t_{1})v(t_{1})dt_{1}\right)
\end{align*}
with survival probability $S(T)=L_{T}[-1]=e^{-\int_{0}^{T}g_{1}(t_{1})dt_{1}}$.

The generating functional is directly related to the characteristic
functional of the stochastic process $\xi(t)=\sum_{j=1}^{n}\delta(t-t_{j})$:
\begin{align}
\Phi_{\xi}[u(t)] & =\langle e^{i\sum_{j=1}^{n}u(t_{j})}\rangle=L_{T}[e^{iu(t)}-1].\label{eq:appendix_charfctl_spike_train}
\end{align}
This relation links the distribution functions $n_{r}$ through the
$g_{r}$ to the cumulants of the spike train. For example, the characteristic
functional of a Poisson spike train is
\begin{equation}
\Phi_{\xi}[u(t)]=\exp\left(\int_{0}^{T}g_{1}(t_{1})(e^{iu(t_{1})}-1)dt_{1}\right).\label{eq:appendix_charfctl_Poisson}
\end{equation}
 Expanding the exponent on the r.h.s.\  of \prettyref{eq:appendix_charfctl_spike_train}
to second order in $u(t)$, we obtain the relations 
\begin{align*}
k_{1}(t_{1}) & =g_{1}(t_{1}),\\
k_{2}(t_{1},t_{2}) & =g_{1}(t_{1})\delta(t_{1}-t_{2})+g_{2}(t_{1},t_{2})
\end{align*}
between the $g_{r}$ and the first two cumulants of the spike train.

\subsection{Gaussian integrals}

We solve several Gaussian integrals using the impressive table by
Owen \citep{Owen80_389}. First, we introduce his notation
\begin{align*}
G(x) & =\frac{1}{2}(1+\erf(x/\sqrt{2}))\\
g(x) & =\frac{1}{\sqrt{2\pi}}e^{-x^{2}/2}
\end{align*}
for the standard normal CDF and PDF. Furthermore, we need Owen's T
function
\begin{align*}
T(h,a) & =\frac{1}{2\pi}\int_{0}^{a}\frac{e^{-\frac{1}{2}h^{2}(1+x^{2})}}{1+x^{2}}dx.
\end{align*}
All formulas were numerically validated using numerical integration
routines implemented in SciPy \citep{Virtanen20_261}.

\subsubsection{GLM error function\label{Appendix B.1}}

Here, we derive \prettyref{eq:GLM_erf_mean} and \prettyref{eq:GLM_erf_corr}.
In the notation of \prettyref{eq:GLM_erf}, we have $\phi(x)=G(x)$.

For the mean, we need the expectation $\langle\phi(z)\rangle$ where
$z$ is Gaussian with mean $\mu$ and variance $\sigma^{2}$. Equivalently,
we can calculate $\langle\phi(\mu+\sigma x)\rangle$ where $x$ is
standard normal. Expressing the standard normal Gaussian expectations
using $g(x)$, we have
\begin{align*}
\langle\phi\rangle & =\int_{-\infty}^{\infty}g(x)G(\mu+\sigma x)dx.
\end{align*}
Using Eq.~(10,010.8) from \citep{Owen80_389}, we get
\begin{align*}
\langle\phi\rangle & =G\left(\frac{\mu}{\sqrt{1+\sigma^{2}}}\right).
\end{align*}
\prettyref{eq:GLM_erf_mean} follows directly.

For the second moment, we need $\langle\phi(z_{1})\phi(z_{2})\rangle$
were $z_{1}$ and $z_{2}$ are jointly Gaussian with mean $\mu$,
variance $\sigma^{2}$ and correlation coefficient $\rho$. Equivalently,
we can calculate $\langle\phi(\mu+\beta x-\alpha y)\phi(\mu+\beta x+\alpha y)\rangle$
where $x$ and $y$ are standard normal and $\alpha=\sigma\sqrt{(1-\rho)/2}$,
$\beta=\sigma\sqrt{(1+\rho)/2}$. Again using $g(x)$ to express the
standard normal Gaussian expectations, we get
\begin{align*}
\langle\phi\phi\rangle & =\int_{-\infty}^{\infty}g(x)I(x)dx\qquad\text{with}\\
I(x) & =\int_{-\infty}^{\infty}g(y)G\left(\mu+\beta x-\alpha y\right)G\left(\mu+\beta x+\alpha y\right)dy.
\end{align*}
Now, we use Eq.~(20,010.3) in \citep{Owen80_389} for $I(x)$ to
obtain
\begin{align*}
\langle\phi\phi\rangle & =\int_{-\infty}^{\infty}g(x)\left(G\left(a+bx\right)-2T\left(a+bx,c\right)\right)dx
\end{align*}
with $a=\mu/\sqrt{1+\sigma^{2}(1-\rho)/2}$, $b=\sigma\sqrt{1+\rho}/\sqrt{2+\sigma^{2}(1-\rho)}$,
$c=\sqrt{1+\sigma^{2}(1-\rho)}$, and Owen's T function $T(h,a)$.
For the final integral, we use Eqs.~(10,010.8) and (c00,010.1) from
\citep{Owen80_389} to derive
\begin{align*}
\langle\phi\phi\rangle & =G\left(\frac{\mu}{\sqrt{1+\sigma^{2}}}\right)-2T\left(\frac{\mu}{\sqrt{1+\sigma^{2}}},\sqrt{\frac{1+\sigma^{2}(1-\rho)}{1+\sigma^{2}(1+\rho)}}\right).
\end{align*}
\prettyref{eq:GLM_erf_corr} follows after subtracting $\langle\phi\rangle^{2}$.

\subsubsection{Free upcrossing probabilities\label{Appendix B.2}}

For the free two-point upcrossing probability, we need integrals of
the form
\begin{align*}
I_{n}(a,b) & =\int_{0}^{\infty}x^{n}g(x)G(ax+b)dx.
\end{align*}
For arbitrary $n$, Eq.~(10,01n.4) from \citep{Owen80_389} provides
the solution
\begin{align*}
I_{n}(a,b) & =\frac{\Gamma((n+1)/2)2^{(n-1)/2}}{\sqrt{2\pi}}F_{n+1,-b}(\sqrt{n+1}a)
\end{align*}
where $F_{\nu,\mu}(x)$ denotes the cumulative distribution function
of noncentral t-distribution with $\nu$ degrees of freedom and noncentrality
parameter $\mu$. Analytical expressions for $F_{\nu,\mu}(x)$ in
terms of $g(x)$, $G(x)$, and $T(h,a)$ can be found in \citep{Owen65_437}
(the ones in \citep{Owen80_389} contain typos). Using these expressions,
the solutions for $n=0,1,2$ are
\begin{align*}
I_{0}(a,b) & =\frac{1}{2}G\left(bB\right)+T\left(bB,a\right),\\
I_{1}(a,b) & =\frac{1}{\sqrt{2\pi}}G(b)+M_{0}(a,b),\\
I_{2}(a,b) & =I_{0}(a,b)+M_{1}(a,b)
\end{align*}
where we used the shorthand notation $B=1/\sqrt{1+a^{2}}$ and
\begin{align*}
M_{0}(a,b) & =aB\,g(bB)\,G(-abB),\\
M_{1}(a,b) & =B^{2}(-abM_{0}(a,b)+ag(b)/\sqrt{2\pi}).
\end{align*}
Since we consider only up to $n=2$, we are spared the increasingly
cumbersome expressions for $n>2$.

\subsection{Free upcrossing probabilities\label{Appendix C}}

The dynamics of the free membrane voltage and the current for the
LIF neuron model are given by
\begin{align}
\dot{U} & =-U+I,\label{eq:appendix_U}\\
\tau_{\mathrm{{s}}}\dot{I} & =-I+\eta,\label{eq:appendix_I}
\end{align}
where we measure time in units of the membrane time constant $\tau_{\mathrm{{m}}}$,
i.e.~we set $\tau_{\mathrm{{m}}}=1$. Furthermore, we set $\langle\eta\rangle=0$,
i.e.~we measure $U$ and $I$ relative to the mean input. Lastly,
we define $t=0$ to be the end of the refractory period, i.e.~the
time when the free dynamics start evolving.

First, we need the distribution of the voltage and the current. Since
$\eta$ is a Gaussian process, both are Gaussian for arbitrary time
arguments. Thus, it is sufficient to calculate the first two conditional
cumulants. Throughout, we assume a correlation-free preparation \citep{haenggi95_239},
i.e.~we assume that $\eta$ and $I$ are uncorrelated prior to $t=0$.

\subsubsection{Non-stationary mean and variance of $U$ and $I$}

We need the non-stationary mean and variance of $U$ and $I$ to calculate
the free upcrossing probability. For a given initial current and initial
voltage, \prettyref{eq:appendix_U} and \prettyref{eq:appendix_I}
lead to
\begin{align*}
I(t) & =I_{0}e^{-t/\tau_{\mathrm{{s}}}}+\frac{1}{\tau_{\mathrm{{s}}}}\int_{0}^{t}e^{-(t-s)/\tau_{\mathrm{{s}}}}\eta(s)ds,\\
U(t) & =U_{0}e^{-t}+\int_{0}^{t}e^{-(t-s)}I(s)ds.
\end{align*}
This leads immediately to the mean
\begin{align*}
\mu_{I}(t) & =I_{0}e^{-t/\tau_{\mathrm{{s}}}},\\
\mu_{U}(t) & =U_{0}e^{-t}+\frac{\tau_{\mathrm{{s}}}}{1-\tau_{\mathrm{{s}}}}I_{0}(e^{-t}-e^{-t/\tau_{\mathrm{{s}}}}).
\end{align*}
To obtain the variances numerically, we use that they follow linear
differential equations: taking the temporal derivatives of $I(t)^{2}$,
$I(t)U(t)$, and $U(t)^{2}$, using \prettyref{eq:appendix_U} and
\prettyref{eq:appendix_I}, and averaging leads to 
\begin{align*}
\frac{\tau_{\mathrm{{s}}}}{2}\dot{\sigma}_{I}^{2} & =-\sigma_{I}^{2}+\sigma_{I\eta}^{2},\\
\tau_{\mathrm{{s}}}\dot{\sigma}_{IU}^{2} & =-(1+\tau_{\mathrm{{s}}})\sigma_{IU}^{2}+\tau_{\mathrm{{s}}}\sigma_{I}^{2}+\sigma_{U\eta}^{2},\\
\frac{1}{2}\dot{\sigma}_{U}^{2} & =-\sigma_{U}^{2}+\sigma_{IU}^{2}.
\end{align*}
The initial conditions for all of the above differential equations
are $\sigma_{I}^{2}(0)=\sigma_{IU}^{2}(0)=\sigma_{U}^{2}(0)=0$. They
are straightforward to solve numerically in the order that they appear,
but they require two additional quantities:
\begin{align*}
\sigma_{I\eta}^{2}(t) & =\frac{1}{\tau_{\mathrm{{s}}}}\int_{0}^{t}e^{-s/\tau_{\mathrm{{s}}}}C_{\eta}(s)ds,\\
\sigma_{U\eta}^{2}(t) & =\frac{1}{1-\tau_{\mathrm{{s}}}}\int_{0}^{t}\left(e^{-s}-e^{-s/\tau_{\mathrm{{s}}}}\right)C_{\eta}(s)ds,
\end{align*}
which can be numerically computed using a composite trapezoidal rule.
If $C_{\eta}(\tau)$ contains a Dirac delta, $C_{\eta}(\tau)=\hat{C}_{\eta}(\tau)+2D\delta(\tau)$,
we have to separate it analytically in $\sigma_{I\eta}^{2}(t)$:
\begin{align*}
\sigma_{I\eta}^{2}(t) & =\hat{\sigma}_{I\eta}^{2}(t)+\frac{D}{\tau_{\mathrm{{s}}}}.
\end{align*}
Note the factor $1/2$ because we only integrate ``half'' of the
Dirac delta. In $\sigma_{U\eta}^{2}(t)$, the Dirac delta does not
contribute because the integrand vanishes at zero, i.e.~$\sigma_{U\eta}^{2}(t)=\hat{\sigma}_{U\eta}^{2}(t)$.

Ultimately, we need the cumulants of $U$ and $\dot{U}$ instead of
$U$ and $I$. To relate the respective quantities, we use \prettyref{eq:appendix_U}.
For the initial conditions, we have
\begin{align*}
\dot{U}_{0} & =I_{0}-U_{0}.
\end{align*}
The first cumulants are
\begin{align*}
\mu_{U}(t) & =U_{0}e^{-t}+(\dot{U}_{0}+U_{0})A(t),\\
\mu_{\dot{U}}(t) & =-\mu_{U}(t)+(\dot{U}_{0}+U_{0})e^{-t/\tau_{\mathrm{{s}}}}\\
 & =-U_{0}e^{-t}+(\dot{U}_{0}+U_{0})B(t),
\end{align*}
where we used \prettyref{eq:appendix_U} for $\mu_{\dot{U}}(t)$ and
abbreviated
\begin{align*}
A(t) & =\frac{\tau_{\mathrm{{s}}}}{1-\tau_{\mathrm{{s}}}}(e^{-t}-e^{-t/\tau_{\mathrm{{s}}}}),\\
B(t) & =e^{-t/\tau_{\mathrm{{s}}}}-A(t).
\end{align*}
The second cumulants do not depend on the initial conditions and we
get from \prettyref{eq:appendix_U}:
\begin{align*}
\sigma_{U\dot{U}}^{2}(t) & =-\sigma_{U}^{2}(t)+\sigma_{IU}^{2}(t),\\
\sigma_{\dot{U}}^{2}(t) & =\sigma_{U}^{2}(t)-2\sigma_{IU}^{2}(t)+\sigma_{I}^{2}(t).
\end{align*}
Finally, we need to marginalize the initial velocity.

We assume that $\dot{U}_{0}$ is Gaussian distributed with mean $\mu_{\dot{U}_{0}}$
and variance $\sigma_{\dot{U}_{0}}^{2}$. Marginalizing $\dot{U}_{0}$
again results in a Gaussian distribution because $p(\dot{U}_{0})$
and $p(U_{1},\dot{U}_{1}\,|\,U_{0},\dot{U}_{0})$ are Gaussian. Hence,
we only need to compute the cumulants. For the mean, we simply have
to replace $\dot{U}_{0}\to\mu_{\dot{U}_{0}}$. The second cumulants
are
\begin{align*}
\tilde{\sigma}_{U}^{2}(t) & =\sigma_{U}^{2}(t)+\sigma_{\dot{U}_{0}}^{2}A(t)^{2},\\
\tilde{\sigma}_{U\dot{U}}^{2}(t) & =\sigma_{U\dot{U}}^{2}(t)+\sigma_{\dot{U}_{0}}^{2}A(t)B(t),\\
\tilde{\sigma}_{\dot{U}}^{2}(t) & =\sigma_{\dot{U}}^{2}(t)+\sigma_{\dot{U}_{0}}^{2}B(t)^{2}.
\end{align*}
With this, we can evaluate the mean and the variance numerically from
the statistics of $\eta(t)$ and $\dot{U}_{0}$.

\subsubsection{Initial velocity distribution}

For the distribution of initial velocities, we assume that the voltage
has reached a stationary distribution by the time it crosses the threshold.
The velocity at an upcrossing of a stationary Gaussian process is
Rayleigh distributed \citep{Stratonovich67}. Because at the threshold
we have $\dot{U}_{\mathrm{up}}=-\theta+I_{\mathrm{up}}$ (remember
that $t=0$ denotes the end of the refractory period, that the membrane
resistance is absorbed into the current, and time is rescaled such
that $\tau_{\mathrm{m}}=1$), the current is also Rayleigh distributed,
\begin{align*}
p(I_{\mathrm{up}}) & =\begin{cases}
\frac{(I_{\mathrm{up}}-\theta)}{\sigma_{I}^{2}}\exp\left(-\frac{(I_{\mathrm{up}}-\theta)^{2}}{2\sigma_{I}^{2}}\right) & \quad\text{for}\quad I_{\mathrm{up}}\ensuremath{\ge\theta}\\
0 & \quad\text{otherwise}
\end{cases}
\end{align*}
where $\sigma_{I}^{2}=-\ddot{C}_{U}(0)$ with the stationary autocorrelation
$C_{U}(\tau)$ of the free voltage. We assume that the further development
of the current is also stationary, and neglect the conditional dependencies
of the transition probability $p(I_{0}\,|\,I_{\mathrm{up}})$ on the
threshold crossing beyond $I_{\mathrm{up}}$, e.g.~on $\dot{I}_{\mathrm{up}}$
and $\ddot{I}_{\mathrm{up}}$. This transition probability can thus
be obtained from the unconstrained (`free') stationary statistics
of the current---not conditioned on a threshold crossing---which
are Gaussian: $p(I_{0}\,|\,I_{\mathrm{up}})=p_{\mathrm{free}}(I_{0},I_{\mathrm{up}})/p_{\mathrm{free}}(I_{\mathrm{up}})$.
The unconstrained joint and instantaneous distributions here function
as auxiliary quantities for computing $p(I_{0}\,|\,I_{\mathrm{up}})$.
The unconstrained joint distribution is a Gaussian with variance $\sigma_{I}^{2}$
and covariance $\sigma_{I}^{2}R_{I}(\tau_{\mathrm{ref}})$ where $R_{I}(\tau)=-\ddot{C}_{U}(\tau)/\sigma_{U}^{2}$.
We derive $C_{U}(\tau)$ most conveniently by Fourier transforming
\prettyref{eq:appendix_U} and \prettyref{eq:appendix_I}, which leads
to $S_{U}(f)=S_{\eta}(f)/(1+(2\pi f)^{2})/(1+(2\pi\tau_{\mathrm{s}}f)^{2})$,
and using the Wiener-Khinchin theorem to obtain the autocorrelation.
From the unconstrained joint and instantaneous distributions, we
obtain the transition probability $p(I_{0}\,|\,I_{\mathrm{up}})$,
which is again a Gaussian with \citep{Williams06}
\begin{align*}
\tilde{\mu}_{I}(\tau_{\mathrm{ref}}) & =I_{\mathrm{up}}R_{I}(\tau_{\mathrm{ref}}),\\
\tilde{\sigma}_{I}^{2}(\tau_{\mathrm{ref}}) & =\sigma_{I}^{2}(1-R_{I}(\tau_{\mathrm{ref}})^{2}).
\end{align*}
Combining this with the Rayleigh-distributed $p(I_{\mathrm{up}})$
yields
\begin{align*}
p(I_{0}) & =\int_{\theta}^{\infty}p(I_{0}\,|\,I_{\mathrm{up}})p(I_{\mathrm{up}})dI_{\mathrm{up}},
\end{align*}
which is not a Gaussian anymore. We only calculate the first two cumulants,
\begin{align*}
\hat{\mu}_{I}(\tau_{\mathrm{ref}}) & =\langle\tilde{\mu}_{I}(\tau_{\mathrm{ref}})\rangle_{I_{0}}=\left(\sqrt{\frac{\pi}{2}\sigma_{I}^{2}}+\theta\right)R_{I}(\tau_{\mathrm{ref}}),\\
\hat{\sigma}_{I}^{2}(\tau_{\mathrm{ref}}) & =\tilde{\sigma}_{I}^{2}(\tau_{\mathrm{ref}})+\langle(\tilde{\mu}_{I}(\tau_{\mathrm{ref}})-\langle\tilde{\mu}_{I}(\tau_{\mathrm{ref}})\rangle_{I_{0}})^{2}\rangle_{I_{0}}\\
 & =\tilde{\sigma}_{I}^{2}(\tau_{\mathrm{ref}})+\frac{4-\pi}{2}\sigma_{I}^{2}R_{I}(\tau_{\mathrm{ref}})^{2},
\end{align*}
and neglect the higher cumulants to arrive at a Gaussian approximation.
Finally, after the refractory time we have $\dot{U}_{0}=-V_{\mathrm{{r}}}+I_{0}$.
Combining the above equations leads to
\begin{align}
\mu_{\dot{U}_{0}} & =\left(\sqrt{\frac{\pi}{2}\sigma_{I}^{2}}+\theta\right)R_{I}(\tau_{\mathrm{ref}})-V_{\mathrm{{r}}},\label{eq:appendix_initial_vel_mean}\\
\sigma_{\dot{U}_{0}}^{2} & =\sigma_{I}^{2}\left(1-\frac{\pi-2}{2}R_{I}(\tau_{\mathrm{ref}})^{2}\right),\label{eq:appendix_initial_vel_var}
\end{align}
which determine the Gaussian approximation of the initial velocity
distribution.

\subsubsection{One-point upcrossing probability}

Here, we calculate the upcrossing probability \prettyref{eq:LIF_n1},

\begin{align*}
n_{1}(t) & =\int_{0}^{\infty}\dot{U}_{1}p(\theta,\dot{U}_{1}\,|\,V_{\mathrm{{r}}})d\dot{U}_{1}.
\end{align*}
Due to the linearity of \prettyref{eq:appendix_U} and \prettyref{eq:appendix_I},
the distribution $p(\theta,\dot{U}_{1}\,|\,V_{\mathrm{{r}}})$ is
a Gaussian with the cumulants we calculated above \citep{haenggi95_239}.
Hence, it takes the form
\begin{align*}
p(\theta,\dot{U}_{1}\,|\,V_{\mathrm{{r}}}) & =\frac{1}{\sqrt{\det(2\pi\mathbf{C})}}\exp\left(-\frac{1}{2}(\mathbf{u}-\bm{\mu})^{T}\mathbf{C}^{-1}(\mathbf{u}-\bm{\mu})\right)
\end{align*}
where $\mathbf{u}^{T}=(\theta,\dot{U}_{1})$ and the mean and the
correlation matrix are given by
\begin{align*}
\bm{\mu} & =\begin{pmatrix}\tilde{\mu}_{U}(t)\\
\tilde{\mu}_{\dot{U}}(t)
\end{pmatrix},\qquad\mathbf{C}=\begin{pmatrix}\tilde{\sigma}_{U}^{2}(t) & \tilde{\sigma}_{U\dot{U}}^{2}(t)\\
\tilde{\sigma}_{U\dot{U}}^{2}(t) & \tilde{\sigma}_{\dot{U}}^{2}(t)
\end{pmatrix}.
\end{align*}
Inverting $\mathbf{C}$ leads to
\begin{align*}
\mathbf{C}^{-1} & =\frac{1}{\det(\mathbf{C})}\begin{pmatrix}\tilde{\sigma}_{\dot{U}}^{2}(t) & -\tilde{\sigma}_{U\dot{U}}^{2}(t)\\
-\tilde{\sigma}_{U\dot{U}}^{2}(t) & \tilde{\sigma}_{U}^{2}(t)
\end{pmatrix},\\
\det(\mathbf{C}) & =\tilde{\sigma}_{U}^{2}(t)\tilde{\sigma}_{\dot{U}}^{2}(t)-\tilde{\sigma}_{U\dot{U}}^{4}(t).
\end{align*}
The exponent of $p(\theta,\dot{U}_{1}\,|\,V_{\mathrm{{r}}})$ takes
the form
\begin{align*}
(\mathbf{u}-\bm{\mu})^{T}\mathbf{C}^{-1}(\mathbf{u}-\bm{\mu}) & =\frac{1}{\det(\mathbf{C})}[a\dot{U}_{1}^{2}-2b\dot{U}_{1}+c^{2}]
\end{align*}
with $a=\tilde{\sigma}_{U}^{2}(t)$, $b=\tilde{\mu}_{\dot{U}}(t)\tilde{\sigma}_{U}^{2}(t)+(\theta-\tilde{\mu}_{U}(t))\tilde{\sigma}_{U\dot{U}}^{2}(t)$
and $c^{2}=\tilde{\mu}_{\dot{U}}(t)^{2}\tilde{\sigma}_{U}^{2}(t)+2(\theta-\tilde{\mu}_{U}(t))\tilde{\mu}_{\dot{U}}(t)\tilde{\sigma}_{U\dot{U}}^{2}(t)+(\theta-\tilde{\mu}_{U}(t))^{2}\tilde{\sigma}_{\dot{U}}^{2}(t)$.

Putting it together, $n_{1}$ is given by
\begin{align*}
n_{1}(t) & =\frac{1}{\sqrt{\det(2\pi\mathbf{C})}}\int_{0}^{\infty}\dot{U}_{1}\exp\left(-\frac{a\dot{U}_{1}^{2}-2b\dot{U}_{1}+c^{2}}{2\det(\mathbf{C})}\right)d\dot{U}_{1}.
\end{align*}
The integral can be solved in terms of an error function:
\begin{align*}
\int_{0}^{\infty}\dot{U}_{1}e^{-\frac{a\dot{U}_{1}^{2}-2b\dot{U}_{1}+c^{2}}{2\det(\mathbf{C})}}d\dot{U} & =\frac{\det(\mathbf{C})}{a}e^{-\tilde{c}^{2}}\\
 & +\frac{\det(\mathbf{C})}{a}e^{-\tilde{c}^{2}}\sqrt{\pi}\tilde{b}e^{\tilde{b}^{2}}\left(1+\erf(\tilde{b})\right)
\end{align*}
where $\tilde{b}=b/\sqrt{2a\det(\mathbf{C})}$ and $\tilde{c}=c/\sqrt{2\det(\mathbf{C})}$.
Thus, we get
\begin{align}
n_{1}(t) & =\frac{\sqrt{\det(\mathbf{C})}}{2\pi\tilde{\sigma}_{U}^{2}(t)}e^{-\tilde{c}^{2}}\left(1+\sqrt{\pi}\tilde{b}e^{\tilde{b}^{2}}\left(1+\erf(\tilde{b})\right)\right)\label{eq:appendix_n1_ana}
\end{align}
for the free upcrossing rate.

\subsubsection{Stationary correlation function of $U$ and $\dot{U}$}

For the stationary two-point upcrossing probability, we need the stationary
correlation functions of $U$, $\dot{U}$, and between $U$ and $\dot{U}$.
The power spectrum of $U$ follows from the power spectrum of $\eta$
using
\[
S_{U}(f)=\frac{S_{\eta}(f)}{(1+(2\pi f)^{2})(1+(2\pi f\tau_{\mathrm{{s}}})^{2})}.
\]
An inverse Fourier transform leads to the stationary correlation function
$C_{U}(\tau)$. For stationary processes, the formulas
\[
C_{U\dot{U}}(\tau)=-C_{\dot{U}U}(\tau)=\dot{C}_{U}(\tau),\qquad C_{\dot{U}}(\tau)=-\ddot{C}_{U}(\tau)
\]
yield the remaining correlation functions. The first formula follows
from $\langle U(t)\dot{U}(t+\tau)\rangle=\frac{d}{d\tau}\langle U(t)U(t+\tau)\rangle$
and $\langle\dot{U}(t)U(t+\tau)\rangle=\langle\dot{U}(t-\tau)U(t)\rangle=-\frac{d}{d\tau}\langle U(t-\tau)U(t)\rangle$,
the second from $\langle\dot{U}(t)\dot{U}(t+\tau)\rangle=\frac{d}{d\tau}\langle\dot{U}(t)U(t+\tau)\rangle=\frac{d}{d\tau}\langle\dot{U}(t-\tau)U(t)\rangle=-\frac{d^{2}}{d\tau^{2}}\langle U(t-\tau)U(t)\rangle$.

\subsubsection{Stationary two-point upcrossing probability}

Here, we calculate the stationary two point upcrossing probability
\prettyref{eq:LIF_n2_stationary},
\begin{align*}
n_{2}(\tau) & =\int_{0}^{\infty}\int_{0}^{\infty}\dot{U}_{2}\dot{U}_{1}p(\theta,\dot{U}_{2};\theta,\dot{U}_{1})d\dot{U}_{1}d\dot{U}_{2}.
\end{align*}
The joint density $p(U_{2},\dot{U}_{2};U_{1},\dot{U}_{1})$ takes
the form
\begin{align*}
p(U_{2},\dot{U}_{2};U_{1},\dot{U}_{1}) & =\frac{1}{\sqrt{\det(2\pi\sigma_{U}^{2}\mathbf{C})}}\exp\left(-\frac{1}{2\sigma_{U}^{2}}\mathbf{u}^{T}\mathbf{C}^{-1}\mathbf{u}\right)
\end{align*}
where $\mathbf{u}^{T}=(U_{1},\dot{U}_{1},U_{2},\dot{U}_{2})$ and
$\sigma_{U}^{2}=C_{U}(0)$. The correlation matrix is given by
\begin{align*}
\mathbf{C} & =\begin{pmatrix}1 & 0 & R(\tau) & \dot{R}(\tau)\\
0 & -\ddot{R}(0) & -\dot{R}(\tau) & -\ddot{R}(\tau)\\
R(\tau) & -\dot{R}(\tau) & 1 & 0\\
\dot{R}(\tau) & -\ddot{R}(\tau) & 0 & -\ddot{R}(0)
\end{pmatrix}
\end{align*}
where we introduced $C_{U}(\tau)=\sigma_{U}^{2}R(\tau)$ and used
$\dot{C}_{U}(0)=0$ for stationary processes with a differentiable
correlation function. Inverting $\mathbf{C}$ is cumbersome and eventually
leads to
\begin{align*}
\mathbf{C}^{-1} & =\frac{1}{\det(\mathbf{C})}\begin{pmatrix}\alpha & \beta & \gamma & \delta\\
\beta & \epsilon & -\delta & \zeta\\
\gamma & -\delta & \alpha & -\beta\\
\delta & \zeta & -\beta & \epsilon
\end{pmatrix}\quad\text{with}\\
\alpha & =\ddot{R}(0)^{2}+\dot{R}(\tau)^{2}\ddot{R}(0)-\ddot{R}(\tau)^{2},\\
\beta & =R(\tau)\dot{R}(\tau)\ddot{R}(0)-\dot{R}(\tau)\ddot{R}(\tau),\\
\gamma & =-R(\tau)\ddot{R}(0)^{2}+R(\tau)\ddot{R}(\tau)^{2}-\dot{R}(\tau)^{2}\ddot{R}(\tau),\\
\delta & =\dot{R}(\tau)\ddot{R}(0)-R(\tau)\dot{R}(\tau)\ddot{R}(\tau)+\dot{R}(\tau)^{3},\\
\epsilon & =-\ddot{R}(0)+R(\tau)^{2}\ddot{R}(0)-\dot{R}(\tau)^{2},\\
\zeta & =\ddot{R}(\tau)-R(\tau)^{2}\ddot{R}(\tau)+R(\tau)\dot{R}(\tau)^{2}.
\end{align*}
The determinant of $\mathbf{C}$ is given by 
\begin{align*}
\det(\mathbf{C}) & =[1-R(\tau)^{2}][\ddot{R}(0)^{2}-\ddot{R}(\tau)^{2}]\\
 & +\dot{R}(\tau)^{2}[2\ddot{R}(0)-2R(\tau)\ddot{R}(\tau)+\dot{R}(\tau)^{2}].
\end{align*}
Now, we have to solve the integrals. The exponent of $p(U_{2},\dot{U}_{2};U_{1},\dot{U}_{1})$
takes the form
\begin{align*}
\mathbf{u}^{T}\mathbf{C}^{-1}\mathbf{u}=\frac{1}{\det(\mathbf{C})}[ & \epsilon(\dot{U}_{1}^{2}+\dot{U}_{2}^{2})+2\zeta\dot{U}_{1}\dot{U}_{2}\\
 & +2(\delta-\beta)\theta(\dot{U}_{2}-\dot{U}_{1})+2(\alpha+\gamma)\theta^{2}].
\end{align*}
With the transformation $v_{1}=\frac{1}{\sqrt{2}}(\dot{U}_{2}-\dot{U}_{1})$
and $v_{2}=\frac{1}{\sqrt{2}}(\dot{U}_{2}+\dot{U}_{1})$, we have
$\dot{U}_{1}^{2}+\dot{U}_{2}^{2}=v_{1}^{2}+v_{2}^{2}$, $\dot{U}_{1}\dot{U}_{2}=\frac{1}{2}(v_{2}^{2}-v_{1}^{2})$
and thus
\begin{align*}
n_{2}(\tau)= & \frac{e^{-\frac{(\alpha+\gamma)\theta^{2}}{\sigma_{U}^{2}\det(\mathbf{C})}}}{2\sqrt{\det(2\pi\sigma_{U}^{2}\mathbf{C})}}\int_{0}^{\infty}e^{-\frac{(\epsilon+\zeta)v_{2}^{2}}{2\sigma_{U}^{2}\det(\mathbf{C})}}\\
 & \times\int_{-v_{2}}^{v_{2}}(v_{2}^{2}-v_{1}^{2})e^{-\frac{(\epsilon-\zeta)v_{1}^{2}+2\sqrt{2}(\delta-\beta)\theta v_{1}}{2\sigma_{U}^{2}\det(\mathbf{C})}}dv_{1}dv_{2}.
\end{align*}
The substitution $\tilde{v}_{i}=v_{i}/\sqrt{2\sigma_{U}^{2}\det(\mathbf{C})}$
simplifies the integrals to
\begin{align*}
n_{2}(\tau)= & \frac{\det(\mathbf{C})^{3/2}}{2\pi^{2}}e^{-\frac{\left((\alpha+\gamma)-\frac{(\beta-\delta)^{2}}{(\epsilon-\zeta)}\right)\theta^{2}}{\sigma_{U}^{2}\det(\mathbf{C})}}\int_{0}^{\infty}e^{-(\epsilon+\zeta)\tilde{v}_{2}^{2}}\\
 & \times\int_{-\tilde{v}_{2}}^{\tilde{v}_{2}}(\tilde{v}_{2}^{2}-\tilde{v}_{1}^{2})e^{-(\epsilon-\zeta)\left(\tilde{v}_{1}-\frac{\beta-\delta}{\epsilon-\zeta}\frac{\theta}{\sqrt{\sigma_{U}^{2}\det(\mathbf{C})}}\right)^{2}}d\tilde{v}_{1}d\tilde{v}_{2}.
\end{align*}
The inner integrals over $\tilde{v}_{1}$ can be solved in terms of
error functions:
\begin{align*}
I_{0}(\tilde{v}_{2};a,b) & \equiv\int_{-\tilde{v}_{2}}^{\tilde{v}_{2}}e^{-a(\tilde{v}_{1}-b)^{2}}d\tilde{v}_{1}\\
 & =\left[\frac{1}{2}\sqrt{\frac{\pi}{a}}\erf(\tilde{v}_{1})\right]_{\sqrt{a}(b-\tilde{v}_{2})}^{\sqrt{a}(b+\tilde{v}_{2})},\\
I_{1}(\tilde{v}_{2};a,b) & \equiv\int_{-\tilde{v}_{2}}^{\tilde{v}_{2}}\tilde{v}_{1}^{2}e^{-a(\tilde{v}_{1}-b)^{2}}dx\\
 & =\left[\frac{1+2ab^{2}}{4a^{3/2}}\sqrt{\pi}\erf(\tilde{v}_{1})\right]_{\sqrt{a}(b-\tilde{v}_{2})}^{\sqrt{a}(b+\tilde{v}_{2})}\\
 & +\left[-\frac{1}{2a^{3/2}}\tilde{v}_{1}e^{-\tilde{v}_{1}^{2}}+\frac{b}{a}e^{-\tilde{v}_{1}^{2}}\right]_{\sqrt{a}(b-\tilde{v}_{2})}^{\sqrt{a}(b+\tilde{v}_{2})},
\end{align*}
where $a=\epsilon-\zeta$ and $b=\frac{\beta-\delta}{\epsilon-\zeta}\frac{\theta}{\sqrt{\sigma_{U}^{2}\det(\mathbf{C})}}$.
Some of the outer integrals over $\tilde{v}_{2}$ can also be solved
in terms of error functions:
\begin{align*}
I_{2}(a,b,c) & \equiv-\frac{b}{a}\int_{0}^{\infty}e^{-c\tilde{v}_{2}^{2}}\left[e^{-\tilde{v}_{1}^{2}}\right]_{\sqrt{a}(b-\tilde{v}_{2})}^{\sqrt{a}(b+\tilde{v}_{2})}d\tilde{v}_{2}\\
 & =\frac{b}{a}\sqrt{\frac{\pi}{a+c}}e^{-ab^{2}+\frac{a^{2}b^{2}}{a+c}}\erf\left(\frac{ab}{\sqrt{a+c}}\right),\\
I_{3}(a,b,c) & \equiv\frac{1}{2a^{3/2}}\int_{0}^{\infty}e^{-c\tilde{v}_{2}^{2}}\left[\tilde{v}_{1}e^{-\tilde{v}_{1}^{2}}\right]_{\sqrt{a}(b-\tilde{v}_{2})}^{\sqrt{a}(b+\tilde{v}_{2})}d\tilde{v}_{2}\\
 & =\frac{1}{2a(a+c)}e^{-ab^{2}}\\
 & -\frac{bc\sqrt{\frac{\pi}{a+c}}}{2a(a+c)}e^{-ab^{2}+\frac{a^{2}b^{2}}{a+c}}\erf\left(\frac{ab}{\sqrt{a+c}}\right),
\end{align*}
with $c=\epsilon+\zeta$. The remaining integrals over $\tilde{v}_{2}$,
i.e.
\begin{align*}
I_{4}(a,b,c) & \equiv-\frac{1+2ab^{2}}{4a^{3/2}}\sqrt{\pi}\int_{0}^{\infty}e^{-c\tilde{v}_{2}^{2}}\left[\erf(\tilde{v}_{1})\right]_{\sqrt{a}(b-\tilde{v}_{2})}^{\sqrt{a}(b+\tilde{v}_{2})}d\tilde{v}_{2},\\
I_{5}(a,b,c) & \equiv\frac{1}{2}\sqrt{\frac{\pi}{a}}\int_{0}^{\infty}\tilde{v}_{2}^{2}e^{-c\tilde{v}_{2}^{2}}\left[\erf(\tilde{v}_{1})\right]_{\sqrt{a}(b-\tilde{v}_{2})}^{\sqrt{a}(b+\tilde{v}_{2})}d\tilde{v}_{2},
\end{align*}
can be solved in terms of Owen's T function $T(h,a)=\frac{1}{2\pi}\int_{0}^{a}\frac{1}{1+x^{2}}e^{-\frac{1}{2}h^{2}(1+x^{2})}dx$
(\citep{Owen80_389}, see Appendix B.2). Combining everything, we
obtain
\begin{align}
n_{2}(\tau) & =\frac{\det(\mathbf{C})^{3/2}}{(2\pi)^{2}ac}I_{\text{ana}}(\tilde{a},\tilde{b},\tilde{c},\tilde{d}),\label{eq:appendix_n2_ana}\\
I_{\text{ana}}(\tilde{a},\tilde{b},\tilde{c},\tilde{d}) & =e^{-\tilde{d}^{2}}+\sqrt{\pi}(1+\tilde{c})\tilde{b}e^{\tilde{b}^{2}-\tilde{d}^{2}}\erf\left(\tilde{b}\right)\nonumber \\
 & +2\pi\sqrt{\tilde{c}}(1/\tilde{c}-2\tilde{a}^{2}-1)e^{\tilde{a}^{2}-\tilde{d}^{2}}T\left(\sqrt{2\tilde{c}}\tilde{b},1/\sqrt{\tilde{c}}\right)\nonumber 
\end{align}
with $\tilde{a}=\sqrt{a}b=\frac{\beta-\delta}{\sqrt{\epsilon-\zeta}}\frac{\theta}{\sqrt{\sigma_{U}^{2}\det(\mathbf{C})}}$,
$\tilde{b}=\frac{a}{\sqrt{a+c}}b=\frac{\beta-\delta}{\sqrt{2\epsilon}}\frac{\theta}{\sqrt{\sigma_{U}^{2}\det(\mathbf{C})}}$,
$\tilde{c}=\frac{c}{a}=\frac{\epsilon+\zeta}{\epsilon-\zeta}$, $\tilde{d}=\sqrt{\alpha+\gamma}\frac{\theta}{\sqrt{\sigma_{U}^{2}\det(\mathbf{C})}}$.
From $n_{2}(\tau)$, we obtain 
\[
Q(\tau)=1-\frac{n_{2}(\tau)}{n_{0}^{2}}\quad\text{and}\quad\eta=2\int_{0}^{\infty}Q(\tau)d\tau
\]
which allow us to evaluate the Stratonovich approximation.

\subsection{Stratonovich approximation\label{Appendix D}}

Here, we compare the full Stratonovich approximation \prettyref{eq:LIF_Stratonovich_approximation},

\begin{align*}
H_{S}(T) & =-\int_{0}^{T}n_{1}(t)\frac{\ln\left(1-\int_{0}^{T}Q(t,t^{\prime})n_{1}(t^{\prime})dt^{\prime}\right)}{\int_{0}^{T}Q(t,t^{\prime})n_{1}(t^{\prime})dt^{\prime}}dt,
\end{align*}
with its approximation \prettyref{eq:LIF_Hertz_rescaled},
\[
h_{S}(t)=\frac{\kappa_{S}}{n_{0}}n_{1}(t),\qquad\kappa_{S}=-\frac{1}{\eta}\ln\left(1-n_{0}\eta\right).
\]
Importantly, both lead to equivalent hazard functions for infinite
times \citep{Stratonovich67}.

To see this, we need two properties of $n_{1}$ and $n_{2}$. First,
the upcrossing probability saturates at a finite value once the transient
effect of the voltage reset is over, $n_{0}=\lim_{t\to\infty}n_{1}(t)$.
Second, $Q(t_{1},t_{2})=1-\frac{n_{2}(t_{1},t_{2})}{n_{1}(t_{1})n_{1}(t_{2})}$
decays to zero for $|t_{2}-t_{1}|\to\infty$ because the upcrossings
decorrelate, $n_{2}(t_{1},t_{2})\to n_{1}(t_{1})n_{1}(t_{2})$. Thus,
one can approximate $\int_{0}^{T}Q(t,t^{\prime})n_{1}(t^{\prime})dt^{\prime}\approx n_{0}\int_{0}^{\infty}Q(t,t^{\prime})dt^{\prime}\equiv n_{0}\eta$
for $0\ll t\ll T$. Then, neglecting the contributions of $\int_{0}^{T}Q(t,t^{\prime})n_{1}(t^{\prime})dt^{\prime}-n_{0}\eta$
for $t$ close to $0$ or $T$ leads to $H_{S}(T)\approx\int_{0}^{T}\frac{\kappa_{S}}{n_{0}}n_{1}(t)dt$.
Neglecting these contributions is justified for large $T$ because
the integral is dominated by the contributions in between these boundaries.
Hence, we arrive at $\lim_{T\to\infty}\frac{d}{dT}H_{S}(T)=\lim_{t\to\infty}h_{S}(t)=\kappa_{S}$
or, in terms of the ISI distribution, $p(T)\sim\exp(-\kappa_{S}T)$
for $t\to\infty$.

\begin{figure}
\includegraphics{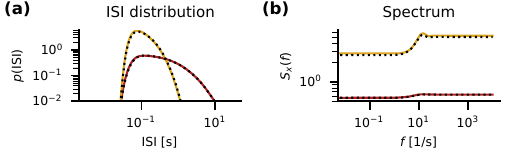}

\caption{Comparison of the full Stratonovich approximation with its approximation
\prettyref{eq:LIF_Hertz_rescaled}. \textbf{(a)} ISI distribution
from Stratonovich approximation (colors) and \prettyref{eq:LIF_Hertz_rescaled}
(black). \textbf{(b)} Same for the power spectra. Parameters as in
\prettyref{fig:LIF_BRN_forward}\textbf{g},\textbf{h}. \label{fig:appendix_Stratonovich_approx}}
\end{figure}

Since the long-time asymptotics are the same, differences can only
occur at short times. In \prettyref{fig:appendix_Stratonovich_approx},
we compare the full Stratonovich approximation with \prettyref{eq:LIF_Hertz_rescaled}
for two representative examples. Fortunately, both the resulting ISI
distributions (\prettyref{fig:appendix_Stratonovich_approx}\textbf{a})
and the power spectra (\prettyref{fig:appendix_Stratonovich_approx}\textbf{b})
agree closely for all times. Solving the full Stratonovich is numerically
challenging (see below); thus, we use the simpler and more efficient
approximation throughout in the main text.

\subsubsection{Numerics}

Here, we develop a numerical implementation of the Stratonovich approximation
that is feasible for long time intervals without excessive demands
on the working memory.

For stationary $Q(t,t^{\prime})=Q(|t^{\prime}-t|)$, the Stratonovich
approximation \prettyref{eq:LIF_Stratonovich_approximation} reads

\begin{align*}
H_{S}(T) & =-\int_{0}^{T}n_{1}(t)\frac{\ln[1-\int_{0}^{T}Q(|t^{\prime}-t|)n_{1}(t^{\prime})dt^{\prime}]}{\int_{0}^{T}Q(|t^{\prime}-t|)n_{1}(t^{\prime})dt^{\prime}}dt.
\end{align*}
With the definition
\begin{align*}
f(T,t) & =\int_{0}^{T}Q(|t^{\prime}-t|)n_{1}(t^{\prime})dt^{\prime}
\end{align*}
we have
\begin{align*}
H_{S}(T) & =-\int_{0}^{T}n_{1}(t)\frac{\ln[1-f(T,t)]}{f(T,t)}dt.
\end{align*}
Since $Q(\tau\to\infty)\to0$, i.e.~it vanishes for long time lags,
we can introduce an associated timescale: $Q(\tau)\approx0$ for all
$\tau>\tau_{Q}$. Similarly, $n_{1}(t\to\infty)\to n_{0}$ on the
timescale $\tau_{n}$ such that $n_{1}(t)\approx n_{0}$ for all $t>\tau_{n}$.

The main problem in computing $H_{S}(T)$ is that a large three-dimensional
grid is necessary for the three time arguments $t$, $t^{\prime}$,
and $T$. To circumvent this problem, we split the domain of integration
such that the full grid is only needed in small subdomains. In the
remainder of the domain, the integrals can be solved by successive
one-dimensional integration.

We consider $f(T,t)$ first. Because $Q(|t^{\prime}-t|)$ vanishes
for $|t^{\prime}-t|>\tau_{Q}$, we know that the integrand only contributes
in the vicinity of $t$. Thus, we can extend the upper limit to infinity,
$f(T,t)\approx f(\infty,t)$ if $t<T-\tau_{Q}$. Accordingly, we split
the integral where possible:
\begin{align*}
H_{S}^{T\le\tau_{Q}}(T)= & -\int_{0}^{T}n_{1}(t)\frac{\ln[1-f(T,t)]}{f(T,t)}dt,\\
H_{S}^{T>\tau_{Q}}(T)\approx & -\int_{0}^{T-\tau_{Q}}n_{1}(t)\frac{\ln[1-f(\infty,t)]}{f(\infty,t)}dt\\
 & +R^{T>\tau_{Q}}(T),\\
R^{T>\tau_{Q}}(T)= & -\int_{T-\tau_{Q}}^{T}n_{1}(t)\frac{\ln[1-f(T,t)]}{f(T,t)}dt.
\end{align*}
The remainder $R^{T>\tau_{Q}}(T)$ becomes constant for $T>\tau_{n}+2\tau_{Q}$
because $n_{1}(t)\approx n_{0}$ in both integrals in this regime
and we can set $R^{T>\tau_{n}+2\tau_{Q}}(T)\approx R^{T>\tau_{n}+2\tau_{Q}}(\tau_{n}+2\tau_{Q})$.
Hence, we only have to calculate the full integral for $H_{S}^{T\le\tau_{Q}}(T)$
and for $R^{T>\tau_{Q}}(T)$ until it is constant.

The remaining integrals in $H_{S}^{T>\tau_{Q}}(T)$ can be solved
successively. First, we solve the convolution integral
\[
f(\infty,t)=\int_{0}^{\infty}Q(|t^{\prime}-t|)n_{1}(t^{\prime})dt^{\prime}
\]
using Fourier transformation. Then, we can insert the result in $H_{S}^{T>\tau_{Q}}(T)$
and solve the integral over $t$. All integrals are approximated by
their respective Riemann sum.

\end{document}